\documentclass[twocolumn,showpacs,aps,prd,amssymb,nofootinbib,nobibnotes,floatfix,longbibliography]{aastex631}

\usepackage{newtxtext,newtxmath}
\usepackage{ae,aecompl}
\usepackage[mathscr]{euscript}

\usepackage[T1]{fontenc}
\usepackage{hyperref}
\usepackage{amsfonts}
\usepackage{amsmath}
\usepackage{mathrsfs}
\usepackage{epsfig}
\usepackage{graphicx}
\usepackage{url}
\usepackage{hyperref}
\usepackage{float}
\usepackage{color}
\usepackage[utf8]{inputenc} 
\usepackage{epsf}
\usepackage{comment}
\usepackage{slashed}
\usepackage{footmisc}
\usepackage{lipsum}
\definecolor{linkcolor}{rgb}{0.0,0.3,0.5}
\usepackage[caption=false]{subfig}
\hypersetup{linkcolor=magenta,citecolor=cerulean,filecolor=cyan,urlcolor=magenta}

\usepackage{color}
\definecolor{cerulean}{rgb}{0.0, 0.48, 0.65}

\definecolor{navy}{rgb}{0.2, 0.0, 1.0}

\definecolor{jungle}{rgb}{0.0, 0.5, 0.0}

\usepackage{color}

\begin{document}

\title{Eccentric Features in the Gravitational Wave Phase of Dynamically Formed Black Hole Binaries}

\author[0000-0002-1516-8653]{Kai Hendriks}
\affiliation{Niels Bohr International Academy, Niels Bohr Institute, Blegdamsvej 17, 2100 Copenhagen, Denmark}

\author[0000-0003-4818-3400]{Lorenz Zwick}
\affiliation{Niels Bohr International Academy, Niels Bohr Institute, Blegdamsvej 17, 2100 Copenhagen, Denmark}

\author[0000-0003-0607-8741]{Johan Samsing}
\affiliation{Niels Bohr International Academy, Niels Bohr Institute, Blegdamsvej 17, 2100 Copenhagen, Denmark}

\shorttitle{Eccentric Features in the Gravitational Wave Phase}
\shortauthors{Hendriks et al.}

\date{\today}

\correspondingauthor{Kai Hendriks}
\email{kai.hendriks@nbi.ku.dk}

\begin{abstract}
We study the gravitational wave (GW) phase shift arising from center-of-mass accelerations of
binary black hole mergers formed dynamically in three-body systems, where both the
inner orbit of the merging binary and the outer orbit are eccentric. We provide a
semi-analytical model and several analytical approximations that allow for fast evaluation
of both the temporal evolution and the maximum value of the phase shift. The highest phase
shifts occur when the binary merges close to the pericentre of the outer orbit, and can in
this case be orders-of-magnitude larger compared to the circular limit.
At high outer orbit eccentricities, the orbital curvature leaves distinct imprints onto the
phase shift if the binary passes the outer pericentre during its inspiral. By comparing
with phase shifts measured in numerical chaotic 3-body scatterings, we show that our
model accurately describes the observed phase of dynamically assembled binary systems
in realistic astrophysical scenarios, providing a way to directly determine their
formation channel via single GW observations.
\end{abstract}

\section{Introduction}
A variety of merging binary black holes (BBHs) has been observed through their emission of gravitational waves (GWs) with
LIGO/Virgo/Kagra (LVK) \citep{2023ApJS..267...29A}. Although the BBH masses \citep{2019ApJ...882L..24A, 2020PhRvL.125j1102A},
spins \citep{2019PhRvD.100b3007Z, 2021PDU....3100791G}, and orbital eccentricities \citep{2019ApJ...883..149A, 2021ApJ...921L..31R, 2022NatAs...6..344G, 2023arXiv230803822T}
have been measured, or at least constrained, the origin of these objects and binaries is still a major unsolved question.
Some proposed formation environments include
isolated binary stars \citep{2012ApJ...759...52D, 2013ApJ...779...72D, 2015ApJ...806..263D, 2016ApJ...819..108B,
2016Natur.534..512B, 2017ApJ...836...39S, 2017ApJ...845..173M, 2018ApJ...863....7R, 2018ApJ...862L...3S, 2023MNRAS.524..426I},
dense stellar clusters \citep{2000ApJ...528L..17P, Lee:2010in,
2010MNRAS.402..371B, 2013MNRAS.435.1358T, 2014MNRAS.440.2714B,
2015PhRvL.115e1101R, 2015ApJ...802L..22R, 2016PhRvD..93h4029R, 2016ApJ...824L...8R,
2016ApJ...824L...8R, 2017MNRAS.464L..36A, 2017MNRAS.469.4665P, Samsing18, 2018MNRAS.tmp.2223S, 2019arXiv190711231S, 2021MNRAS.504..910T, 2022MNRAS.511.1362T},
galactic nuclei (GN) \citep{2009MNRAS.395.2127O, 2015MNRAS.448..754H,
2016ApJ...828...77V, 2016ApJ...831..187A, 2016MNRAS.460.3494S, hoa18, 2018ApJ...865....2H,2019ApJ...885..135T, 2019ApJ...883L...7L,2021MNRAS.502.2049L, 2023MNRAS.523.4227A},
active galactic nuclei (AGN) discs \citep{2017ApJ...835..165B,  2017MNRAS.464..946S, 2017arXiv170207818M, 2020ApJ...898...25T, 2022Natur.603..237S, 2023arXiv231213281T, Fabj24},
single-single GW captures of primordial black holes \citep{2016PhRvL.116t1301B, 2016PhRvD..94h4013C,
2016PhRvL.117f1101S, 2016PhRvD..94h3504C},
and very massive stellar mergers \citep{Loeb:2016, Woosley:2016, Janiuk+2017, DOrazioLoeb:2018}. Generally, it has proven difficult to tell these different
channels apart using GWs alone. However, studies have shown that different classes of channels share specific properties, e.g.
dynamically formed BBH mergers will give rise to a significant fraction of eccentric
mergers \citep[e.g.][]{2006ApJ...640..156G, 2014ApJ...784...71S, 2017ApJ...840L..14S, Samsing18a, Samsing2018, Samsing18, 2018ApJ...855..124S,
2018MNRAS.tmp.2223S, 2018PhRvD..98l3005R, 2019ApJ...881...41L,2019ApJ...871...91Z, 2019PhRvD.100d3010S, 2019arXiv190711231S},
which is different from those forming through isolated binary evolution.
The BBH spins \citep[e.g.][]{2000ApJ...541..319K, 2016ApJ...832L...2R,2018ApJ...863...68L}, as well as the mass
distribution \citep[e.g.][]{2017ApJ...846...82Z,2021MNRAS.505.3681S} can also be used to disentangle different formation channels.
While this is encouraging, the group of dynamically formed BBH mergers also consists of several different channels,
with overlapping observed quantities. For example, GW captures forming in GN \citep[e.g.][]{2009MNRAS.395.2127O}, 
hierarchical Lidov-Kozai-triple configurations \citep[e.g.][]{hoa18, 2019ApJ...881...41L}, and 
few-body interactions in globular cluster (GCs) \citep[e.g.][]{Samsing18} and
AGNs \citep[e.g.][]{2022Natur.603..237S, Fabj24} all give rise to eccentric mergers with similar distributions.

\begin{figure*}
    \centering
    \includegraphics[width=0.95\textwidth]{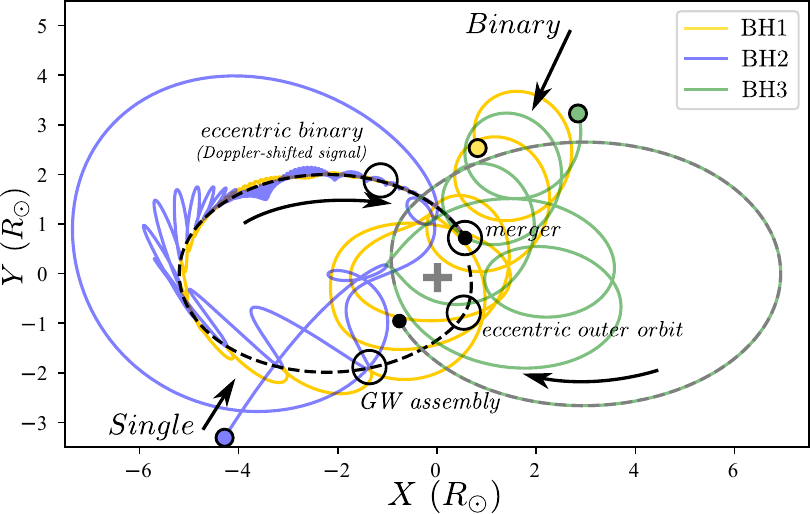}
    \caption{{\bf Formation of a GW phase-shifted source.} Illustrative example of a scattering between a binary black hole (BH1,BH3) and an incoming single black hole (BH2) that results in a highly eccentric BBH merger (BH1,BH2), while all three objects are still bound to each other. The eccentric BBH merger therefore inspirals and merge on an eccentric orbit around the remaining BH3, which gives rise to a Doppler-shifted GW signal due to binary COM acceleration in the COM frame. This GW-shift can be detected, and will leave imprints on how the specific BBH assembly and merger takes place, thereby revealing the BBH origin. This is further illustrated in Fig. \ref{fig:phase_shift}. In this example, we use $m_1=15M_\odot$, $m_2=5M_\odot$, and $m_3=15M_\odot$.}
    \label{fig:scatteringEX1}
\end{figure*}

One very interesting measure of the formation environment is how the presence of a third body induces a phase shift onto the gravitational waveform in the observer frame. In such a scenario, the BBH evolves while its COM is being accelerated. The acceleration causes a time-dependent Doppler shift in the gravitational waveform, which in principle can be measured and used to probe the exact dynamical environment within which the binary lives as it merges. Therefore, looking for GW phase shifts, or more generally environmental effects, could be the key to help us tell apart the large suit of sub-channels, provided they are present in the observable GW bands \citep[e.g.][]{tiwari_accelerated_2024, 2024arXiv240715117T}.

As of now, this scenario has been looked at in the case of circular orbits for both the BBH itself (inner orbit) and its orbit around the perturbing BH (outer orbit) \citep{2017ApJ...834..200M}, and a circular binary on an eccentric outer orbit \citep{2018PhRvD..98f4012R}. Recently, we extended this problem to the case of BBHs formed dynamically \citep{samsing_gravitational_2024}. Here the BBH itself is eccentric and inspirals on a circular orbit around the third object, which naturally gives rise to more unique imprints in the gravitational waveform. In this present paper we take the next step and model the GW phase of a general eccentric BBH inspiralling on an eccentric outer orbit near another BH.

This generalisation is vital as it accurately represents the case for dynamically assembled BBH mergers \citep[see e.g.,][]{2000ApJ...528L..17P,Samsing14,2016PhRvD..93h4029R, 2016ApJ...824L...8R,2017MNRAS.464L..36A,2018MNRAS.tmp.2223S,2018ApJ...863...68L,hoa18,2022MNRAS.511.1362T}. An example is depicted in Fig. \ref{fig:scatteringEX1}, that shows the formation of a BBH inspiral and merger formed through a chaotic scattering process between three black holes. These types of interactions are known to frequently take place in GCs and other dense environments, and have especially been suggested to be among the most reliable ways for forming BBH mergers with residual eccentricity across the observable GW bands \citep{Samsing18}. Moreover, including and considering outer orbit eccentricity is not expected to just add a small perturbation to the problem, as the maximum GW phase shift scales as $\propto R^{-2}$, where $R$ is the distance between the BBH and the third BH \citep{2017ApJ...834..200M,samsing_gravitational_2024}. This implies that if the BBH merges close to the pericentre of its outer orbit, the GW phase shift may increase dramatically compared to a circular outer orbit with the same period. This situation is in fact seen in Fig. \ref{fig:scatteringEX1}, where {\it `merger'} takes place near pericentre of the outer orbit. 

The aim of this paper is to investigate the GW Doppler shift and corresponding GW phase shift of an eccentric BBH on an eccentric orbit around a third BH, by quantifying effects that are unique to the outer orbit being eccentric. Effects for different outer orbit families have recently been
explored using an expansion in acceleration \citep{2024arXiv240715117T}; In our present paper we expand on this, by further developing a semi-analytical method that solves for the full orbit from which we discover several new non-linear effects in the resultant GW phase shift.

The paper is organised as follows. We start in Sec. \ref{sec:theory} by outlining our theory for how we calculate GW phase shifts in the general case where both the inner and outer orbits can be eccentric. This is followed by Sec. \ref{sec:phase_shift}, where we explain the main characteristics and observable features of the eccentric outer orbit. In Sec. \ref{sec:astro} we discuss implications in astrophysically relevant cases, where we highlight phase shifts in scatterings and the impact of tidal forces on the phase shift. We then conclude the study and provide future prospects in Sec. \ref{sec:summary}.

\section{Theory}\label{sec:theory}

\begin{figure}
    \centering
    \includegraphics[width=.48\textwidth]{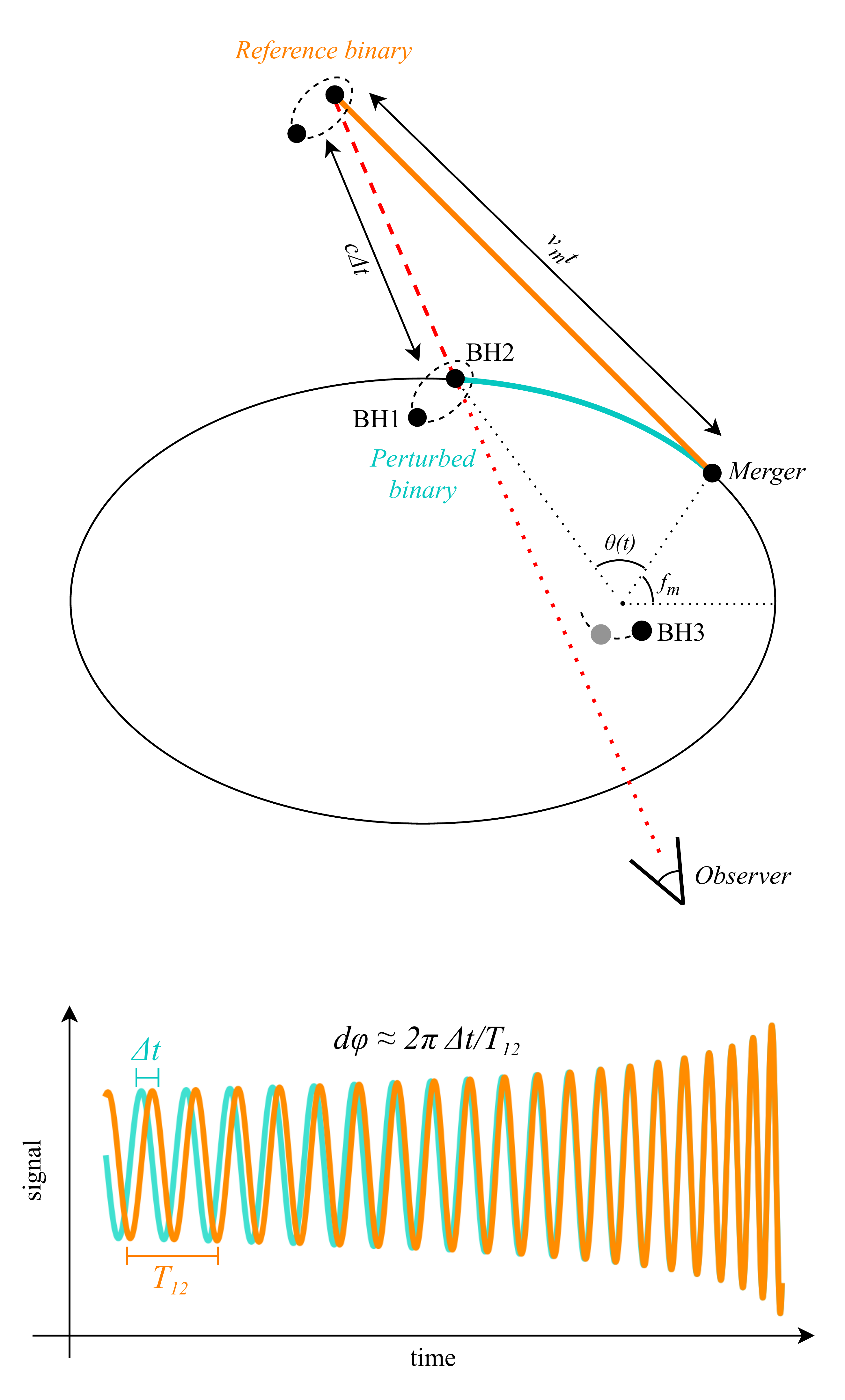}
    \caption{{\bf Illustration of our setup}. {\textit{Top:}} The turquoise curve represents the real trajectory of the COM of an eccentric
    binary (BH1, BH2), whose COM is on an eccentric orbit around a perturber (BH3). In orange, we show the trajectory of the same binary,
    if it had evolved in isolation. The reference binary has a constant speed $v_m$ equal to the orbital speed of the COM of the perturbed binary
    at merger, which happens at true anomaly $f_m$. A Rømer delay $\Delta t$ arises due to the difference in light travel time between the
    two scenarios towards an observer, who is depicted in the bottom right. {\textit{Bottom:}} schematic representation of how the
    Rømer delay is imprinted in the gravitational waveform. The perturbed waveform receives a time-dependent Doppler shift with
    respect to its isolated counterpart, which we can turn into a phase shift by equating it with the orbital period of the binary (BH1, BH2)
    at the same point in time: $d\phi \approx 2\pi \Delta t / T_{12}$.}
    \label{fig:phase_shift}
\end{figure}

As described in \cite{2017ApJ...834..200M, samsing_gravitational_2024}, the GW Doppler shift arising from the acceleration on an outer orbit can equivalently be viewed as
the GW phase shift from Rømer delay (RD) between the binary trajectory and some non-accelerating reference trajectory. In Fig. \ref{fig:phase_shift}
we show our setup in this framework. We depict in turquoise the trajectory of an initially highly (inner-orbit) eccentric BBH (BH1, BH2) that 
inspirals and merges on an eccentric (outer) orbit around BH3 (see also Fig. \ref{fig:scatteringEX1}). Each of these BHs have a mass $m$ denoted
$m_1$, $m_2$ and $m_3$, respectively.
We compare this to a fictional reference binary, shown in orange. This binary has
exactly the same parameters (i.e. initial eccentricity, semi-major axis, time to coalescence, binary masses), but moves on a straight line with a
constant velocity $v_m$ that is equal to the tangential velocity of the perturbed binary at merger. Tracing these two paths back in time from merger,
their gradual deviation is what gives rise to the time-dependent Rømer delay.
If the distance between the two trajectories at some time $t$ is given
by $l(t)$ (red dashed line in Fig. \ref{fig:phase_shift}), then their relative maximum Rømer delay, $\Delta t (t)$, is
\begin{align}
\Delta t (t) = l(t)/c,
\label{eq:def_deltat_RD}
\end{align}
where $c$ is here the speed of light. As shown in the bottom sketch of Fig. \ref{fig:phase_shift}, the
difference in arrival time between the GWs from the two trajectories
induces a phase shift onto the gravitational waveform.

It is important to realise that the $\Delta t$ shown in Fig. \ref{fig:phase_shift} is the maximum time delay between the two
trajectories. In reality, an observer only sees the projection of $\Delta t$ in their own
line-of-sight \citep[e.g.][]{2017ApJ...834..200M, samsing_gravitational_2024}. However, in this work, we solely focus on the maximum possible phase
shift that we denote simply by $d\phi$; factoring in the trivial observer dependent factor will be investigated in later work.
Other sources of dephasing in gravitational waveforms in dynamically assembled systems includes General Relativistic (GR) effects, such as Shapiro delay \citep{Backer1986}, gravitational redshift \citep{2017ApJ...834..200M,2021chen}, and gravitational lensing \citep[e.g.][]{wang1999, takashi2003, ezquiaga2021, lo2024}, as well as special relativistic effects \citep{2019alejandro,2023yan}, aberration \citep{2020alejandro}, and classical tidal effects \citep{samsing_gravitational_2024}. Relevant discussions of similar setups can be found in \citep[e.g.][]{2011PhRvD..83d4030Y,2017PhRvD..96f3014I,
2018PhRvD..98f4012R, 2019PhRvD..99b4025C, 2019ApJ...878...75R, 2019MNRAS.488.5665W,
2020PhRvD.101f3002T, 2020PhRvD.101h3031D, 2021PhRvL.126j1105T, 2022PhRvD.105l4048S, 2023PhRvD.107d3009X, 2023arXiv231016799L, tiwari_accelerated_2024}.

In the sections below we start by deriving our semi-analytical description for the Rømer time delay (Sec. \ref{sec:rømer}), as well as our
expression for the corresponding GW phase shift (Sec. \ref{sec:phaseshift}). Subsequently, in Sec. \ref{sec:numerical} we lay out the
general numerical procedure to obtain these quantities. Finally, we provide analytical approximations to the phase shift and
illustrate how they can be used to estimate the GW phase shift created for outer eccentric orbits when the observational time
window is relatively small (Sec. \ref{sec:approx}).

\subsection{Rømer Time Delay}\label{sec:rømer}

Using Kepler's Laws, one can use geometrical arguments and standard mechanics to derive the Rømer time delay and corresponding phase shift,
as we illustrate in the following. The analytical solution to the case where the inner orbit is eccentric and the
outer orbit is circular was presented in \cite{samsing_gravitational_2024},
and we therefore go right into presenting the solution for the RD in the more general case of an eccentric outer orbit.

For deriving the RD $\Delta{t}$ we start by calculating the distance between the the turquoise perturbed binary
COM and the orange reference binary COM at a given time $t$ (see Fig. \ref{fig:phase_shift}), defined as the time in
the restframe of the 3-body system. The fact that the position and evolution of a Keplerian eccentric orbit cannot be
written in closed form, makes the simple geometrically description slightly more complex compared to the circular case.

In this work we find it convenient to split up the true anomaly $f$ in two parts,
\begin{equation}
f(t) = \theta(t) + f_m,
\label{eq:ft}
\end{equation}
where $f_m$ is the angle at which the binary merges and $\theta(t)$ is the angle between the position vector of
the turquoise binary COM at time $t$ and its position vector at merger (see Fig. \ref{fig:phase_shift}).
Another important quantity to note is the distance between the BBH and the third object, $r(f)$, which is a function of $f$
as given by,
\begin{align}
    r(f) = \frac{a_\mathrm{out}(1-e_\mathrm{out}^2)}{1+e_\mathrm{out}\cos f},
    \label{eq:rf}
\end{align}
where $a_\mathrm{out}$ and $e_\mathrm{out}$ denote the semi-major axis and eccentricity of the BBH outer orbit, respectively.
With this set of equations we are now in a position to calculate the distance $l$ between the reference binary undergoing a
straight line motion with velocity $v_m$, and the perturbed or true binary moving on the eccentric outer orbit.
Using standard geometry the distance $l$ is at time $t$ given by,
\begin{align}
\begin{split}
    l(t) &= \mu \Bigg[\left[r(f_m)\cos(f_m) + v_{m,x}t - r(f)\cos(f)\right]^2 \\
    &\qquad + \left[r(f_m)\sin(f_m) + v_{m,y}t - r(f)\sin(f)\right]^2\Bigg]^{1/2} \\
    &= \mu \Bigg[v_m^2 t^2 \\
    &\qquad - 2t\bigg[v_{m,x}(r(f)\cos(f) - r(f_m)\cos(f_m)) \\ 
    &\qquad\quad + v_{m,y}(r(f)\sin(f) - r(f_m)\sin(f_m)))\bigg] \\
    &\qquad + r^2(f_m) + r^2(f) - 2r(f_m)r(f)\cos(f - f_m)\Bigg]^{1/2},
\end{split}
\label{eq:l}
\end{align}
where the quantities $v_{m,x}$ and $v_{m,y}$ are the $x,y$ velocity components at merger, $f$ depends on $t$ through Eq. \ref{eq:ft},
and $\mu = m_3/m_{123}$, where $m_{123} = m_1 + m_2 + m_3$. In the rest of the paper, we will refer to the quantity $r(f_m)$, i.e.
the distance to the perturber at merger, simply as $r_m$. These quantities relate to the angular and radial velocities of the binary COM at phase $f$,
\begin{align}
    v_\text{ang}(f) = \frac{\sqrt{Gm_{123}a_\text{out}(1-e_\text{out}^2))}}{r(f)},
\end{align}
and
\begin{align}
    v_\text{rad}(f) = e_\text{out}\sin(f) \sqrt{\frac{Gm_{123}a_\text{out}}{(1-e_\text{out}^2)}},
\end{align}
as
\begin{align}
    \Vec{v}(f) = -\left(
    \begin{array}{c}
    v_\text{ang}\sin(f) -v_\text{rad}\cos(f)  \\
    -v_\text{ang}\cos(f) -v_\text{rad}\sin(f)
    \end{array}\right).
\end{align}
Here, we explicitly mention that $\Vec{v}_m = \Vec{v}(f_m)$.

Because the outer orbit is eccentric, we cannot obtain an analytical relation between $f$ and $t$. Therefore, we calculate $l(t)$ numerically
using the equations above. From this calculation of the distance $l$ between the two trajectories, the RD can now be calculated using Eq. \ref{eq:def_deltat_RD}.

\subsection{Gravitational Wave Phase shift}\label{sec:phaseshift}

As shown in \cite{samsing_gravitational_2024}, the maximum GW phase shift at a given time $t$, $d\phi(t)$, evaluated in the observer frame can be approximated by
the maximum RD, $\Delta{t}$, divided by the inner orbital time of the BBH, $T_{12}$, times $2\pi$ to get it in radians,
\begin{align}
    d\phi(t) \approx 2\pi \frac{\Delta t(t)}{T_{12}(t)} = \frac{2\pi}{c} \frac{l(t)}{T_{12}(t)}.
    \label{eq:dphi}
\end{align}
This is the relevant factor to consider in relation to observational constraints \citep[e.g.][]{2017ApJ...834..200M}.

To derive $d\phi$ for outer eccentric orbits we have the RD $\Delta{t}$ from Eq. \ref{eq:l} that can be evaluated at time $t$.
The corresponding orbital time $T_{12}$ can also be defined at time $t$, which follows from Kepler's Law,
\begin{align}
    T_{12} = \frac{2\pi}{\sqrt{G(m_1+m_2)}} a^{3/2},
    \label{eq:T12}
    \end{align}
where $a$, and associated $e$, refer to the semi-major axis and the eccentricity of the merging BBH inner orbit, respectively.
For solving the evolution, from assembly to merger of the BBH due to GW angular momentum and energy losses as it spirals in, we use the
relation for $a(e)$ presented in \cite{peters_gravitational_1964} (which, hereafter, we will refer to as Peters64),
\begin{align}
    a(e) = a_0\frac{g(e)}{g(e_0)},
    \label{eq:a_e}
\end{align}
where
\begin{align}
    g(e) = \frac{e^{12/19}}{1 - e^2}\bigg(1 + \frac{121}{304}e^2\bigg)^{870/2299},
    \label{eq:ge}
\end{align}
and $a_0$ and $e_0$ are the semi-major axis and eccentricity of the binary at some
reference point in their evolution.

Note that in order to calculate the phase shift we need $\Delta t(t)$ and $T_{12}$ in terms of either $t$ or $e$.
While a closed form for $e(t)$ does not exist, an expression for $t(e)$ is given in Peters64. However, this form is somewhat inaccurate for
estimating the correct inspiral time when the binary is highly eccentric. The general problem relates to the fact that the
orbital elements, $a$ and $e$, are defined using Newtonian prescriptions, which break down especially for eccentric BBH inspirals
where these quantities not only are drastically changing as a function of time, but also ill-defined. Instead we here
make use of the fitting function presented in \citep{zwick_improved_2020, zwick_improved_2021}, which provides a more accurate
estimate for the merger time than the Peters64 solution.
The fit takes the form
\begin{align}
\begin{split}
    t_{c} &= \frac{5c^5 (1+q)^2}{256G^3m_{12}^3q}\frac{a(e)^4}{f(e)} 8^{1-\sqrt{1 - e}}
    \exp\left(\frac{2.8r_S}{a(e)(1-e)}\right) \\
    &\quad \times \Bigg\{1 + \bigg[-1 + \exp\left(\frac{2.2r_S}{a(e)(1-e)}\right) \\
    &\qquad\quad + \left(\frac{3.8r_S}{a(e)(1-e)}\right)^{3/2}\bigg](1-e)^2 \\
    &\qquad\quad - \left(\frac{3.8r_S}{a(e)(1-e)}\right)^{3/2}(1-e)\Bigg\},
    \label{eq:tc}
\end{split}
\end{align}
where $m_{12} = m_1+m_2$, $q=m_2/m_1\geq1$, $r_S = {2G(m_1+m_2)}/{c^2}$ and
\begin{align}
    f(e) = \bigg(1 + \frac{73}{24}e^2 + \frac{37}{96}e^4\bigg)(1 - e^2)^{-7/2}.
\end{align}
The relation $a(e)$ from Peters64 given by Eq. \ref{eq:a_e}, has been shown to still provide an accurate description, and we therefore continue using this throughout this work.

\subsection{Numerical Procedure}\label{sec:numerical}

\begin{figure}
    \centering
    \includegraphics[width=.46\textwidth]{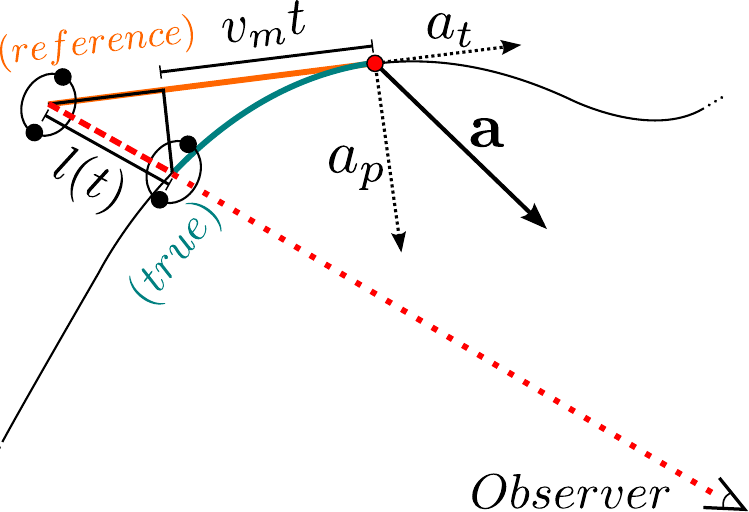}
    \caption{{\bf Binary undergoing accelerated motion.} Schematic depicting a binary with an accelerated centre of mass (turquoise line) with respect to an inertial reference binary (orange line). The presence of curvature in the binary's path creates a time varying Doppler shift that influences its orbital frequency, as measured by an observer along a given line of sight (dotted red line).}
    \label{fig:acc_gen_fig}
\end{figure}

Here we briefly lay out the numerical procedure we follow to calculate the phase shift $d\phi$ in our setup.
Our model requires a total of 8 inputs: 4 from for the inner orbit BBH evolution ($m_1, m_2, a_0, e_0$),
where $a_0, e_0$ here refer to the initial
values right after assembly, and 4 for the outer orbit evolution of the BBH around the perturber ($m_3, f_m, a_\mathrm{out}, e_\mathrm{out}$).
First we derive the inner orbital evolution of the inspiralling BBH.
\begin{itemize}
\item{We first make an array of values for $e$ ranging from $e_0$ all the way to $e = 0$, from which we can make an array for
$a(e)$ using Eq. \ref{eq:a_e}.}
\item{With our derived array $a(e)$ we then calculate the corresponding BBH inner orbital time $T_{12}$, given by Eq. \ref{eq:T12}.}
\item{We then compute the array for the corresponding merger time $t_c$ given by Eq. \ref{eq:tc}.}
\end{itemize}
For estimating the phase shift, we have to couple this procedure with the outer orbital evolution of the BBH around the perturber.
\begin{itemize}
\item{We first choose orbital elements for the (non-decaying) outer orbit, $a_\mathrm{out}$ and $e_\mathrm{out}$, and an angular position for where the merger takes place, $f_m$.}
\item{The computed array of the BBH merger time, $t_c$ (from the inner binary as described above) is then used with $a_\mathrm{out}$, $e_\mathrm{out}$, and $f_m$, to numerically obtain an array for the corresponding true anomaly $f$.}
\item{From this we now have the components to compute the distance $l(t)$, defined in Eq. \ref{eq:l}, from which we find the maximum RD given by Eq. \ref{eq:def_deltat_RD}.}
\item{Finally, we now have matching arrays with $\Delta{t}$ and orbital time $T_{12}$, as a function of either time $t$, orbital eccentricity $e$, or true anomaly $f$. We then calculate the maximum GW phase shift $d\phi$ given by Eq. \ref{eq:dphi}.}
\end{itemize}

\subsection{Analytical Approximations}\label{sec:approx}

Before we move to the main results derived from our procedure outlined above, we first consider a few highly useful and general
approximations to the GW phase shift that allow one to estimate its magnitude analytically and corresponding relevant scalings.
We especially describe how these can be used for fast estimations of the phase shift for eccentric BBH mergers formed during chaotic
few-body interactions, which are being explored both in controlled experiments \citep[e.g.][]{Samsing18, samsing_gravitational_2024} and in MC codes such as CMC \citep{kremer2019} where merger statistics are based on millions of scatterings.

Below we first present an analytical approximate solution to the general case of an inspiralling eccentric BBH moving on an arbitrary trajectory,
while being subject to an arbitrary acceleration vector \citep[for recent work on circular binaries, see also][]{vijaykumar2023, tiwari_accelerated_2024}. We then consider solutions specific to inner and outer
eccentric orbits applicable to PN N-body studies.

\subsubsection{Phase Shift from a General Acceleration}\label{sec:approx_gen}

We start by considering the case of a BBH moving on an arbitrary trajectory, as shown in Fig. \ref{fig:acc_gen_fig}. At the point of merger
the BBH has a velocity $v_m$ and is subject to a general acceleration $\bf a$.
While the velocity vector is tangential to the trajectory of the BBH,
the acceleration vector does not have to be either parallel or perpendicular to the the trajectory. In a circular motion, the acceleration
vector will be perpendicular to the velocity vector, but this is not the case for e.g. a simple eccentric orbit,
where the velocity- and the acceleration vectors only are perpendicular to each other at peri- and apocentre. The question is what the
maximum phase shift $d\phi$ can be for such a general trajectory and acceleration (see also \cite{2024arXiv240715117T}). We here estimate this to linear order,
where we assume the acceleration and velocity is constant near the point of merger.
As shown in \cite{samsing_gravitational_2024},
the maximum value of the phase shift when the BBH inspirals with non-zero inner orbit eccentricity, e.g. without taking into account
the observer position, will arise close to merger, which justifies this assumption.

Using the notation from Fig. \ref{fig:acc_gen_fig}, we first estimate the distance $l$, i.e. $l = c\Delta{t}$ as shown in Fig. \ref{fig:phase_shift},
between the reference trajectory and the real trajectory while taking into account both the tangential- and the perpendicular acceleration components,
\begin{align}
\begin{split}
    l(t)^2 & = \left(a_t t^2/2\right)^2 + \left(a_p t^2/2\right)^2 = {|\bf a|}^2 \left(t^2/2\right)^2,
\end{split}
\end{align}
leading to
\begin{align}
\begin{split}
    l(t) & = \frac{1}{2}{|\bf a|}t^{2}.
\end{split}
\end{align}
Using the general expression for $d\phi$ given by Eq. \ref{eq:dphi}, one sees that $d\phi$ for an inspiralling BBH on an arbitrary orbit subject to a general acceleration vector $\bf a$ near merger is
\begin{equation}
d{\phi} \approx \frac{\pi}{c} \frac{|{\bf a}|t^2}{T_{12}}.    
\label{eq: dphi_gen_a}
\end{equation}
This can easily be evaluated for any system, including N-body systems, which allows for a fast phase shift estimation as we will further explain in Sec. \ref{sec:Fast Estimator Phaseshift} below.

\subsubsection{Circular Approximation}\label{sec:Phaseshift from ecc orbits}

The approximation above is at the linear level. However, when considering the specific case of a BBH moving on an eccentric orbit
it is possible to provide a slightly more accurate, but still fast, estimate by approximating the trajectory by a circle around
the point of merger instead of a line. For this, we continue under the assumption that the BBH is moving on a circle that is
tangent to the point of merger with a radius corresponding to the distance between the BBH and the three-body COM, $\mu r_m$,
and effective velocity $v$ defined as,
\begin{align}
    v = \sqrt{Gm_{123}/r_m}.
\end{align}
By now defining the angle $\theta$ to be the angular evolution of the BBH on that circle,
\begin{align}
    \theta(t) = {v t}/{r_m},
\end{align}
the RD is found in this approximation using standard geometry (see \cite{samsing_gravitational_2024} for further details),
\begin{align}
\begin{split}
    \Delta t(t) &= \frac{2\mu r_m}{c} [(\theta/2)^2 + \sin^2(\theta/2) \\
    &\qquad- 2(\theta/2)\sin(\theta/2)\cos(\theta/2)]^{1/2}.
    \label{eq:dt_circ}
\end{split}
\end{align}
By now substituting this relation into Eq. \ref{eq:dphi}, one finds
\begin{align}
\begin{split}
    d\phi &\approx 2\frac{\sqrt{Gm_{12}}\mu r_m}{a(e)^{3/2}} \bigg[\frac{Gm_{123}}{4r_m^3}t_c(e)^2 + \sin^2\left(\sqrt{\frac{Gm_{123}}{r_m}} \frac{t_c(e)}{2r_m}\right) \\
    &\qquad- \sqrt{\frac{Gm_{123}}{r_m}}\frac{t_c(e)}{r_m}\sin\left(\sqrt{\frac{Gm_{123}}{r_m}} \frac{t_c(e)}{2r_m}\right) \\
    &\qquad\quad \times \cos\left(\sqrt{\frac{Gm_{123}}{r_m}}\frac{t_c(e)}{r_m}\right)\bigg]^{1/2},
    \label{eq:dphi_circ}
\end{split}
\end{align}
where $t_c$ is given by Eq. \ref{eq:tc}. If one uses the eccentricity as evolution factor, one can derive $d\phi(e)$ in closed form for
any value of $e$ as described in the following.

\begin{figure*}
    \centering
    \includegraphics[width=.8\textwidth]{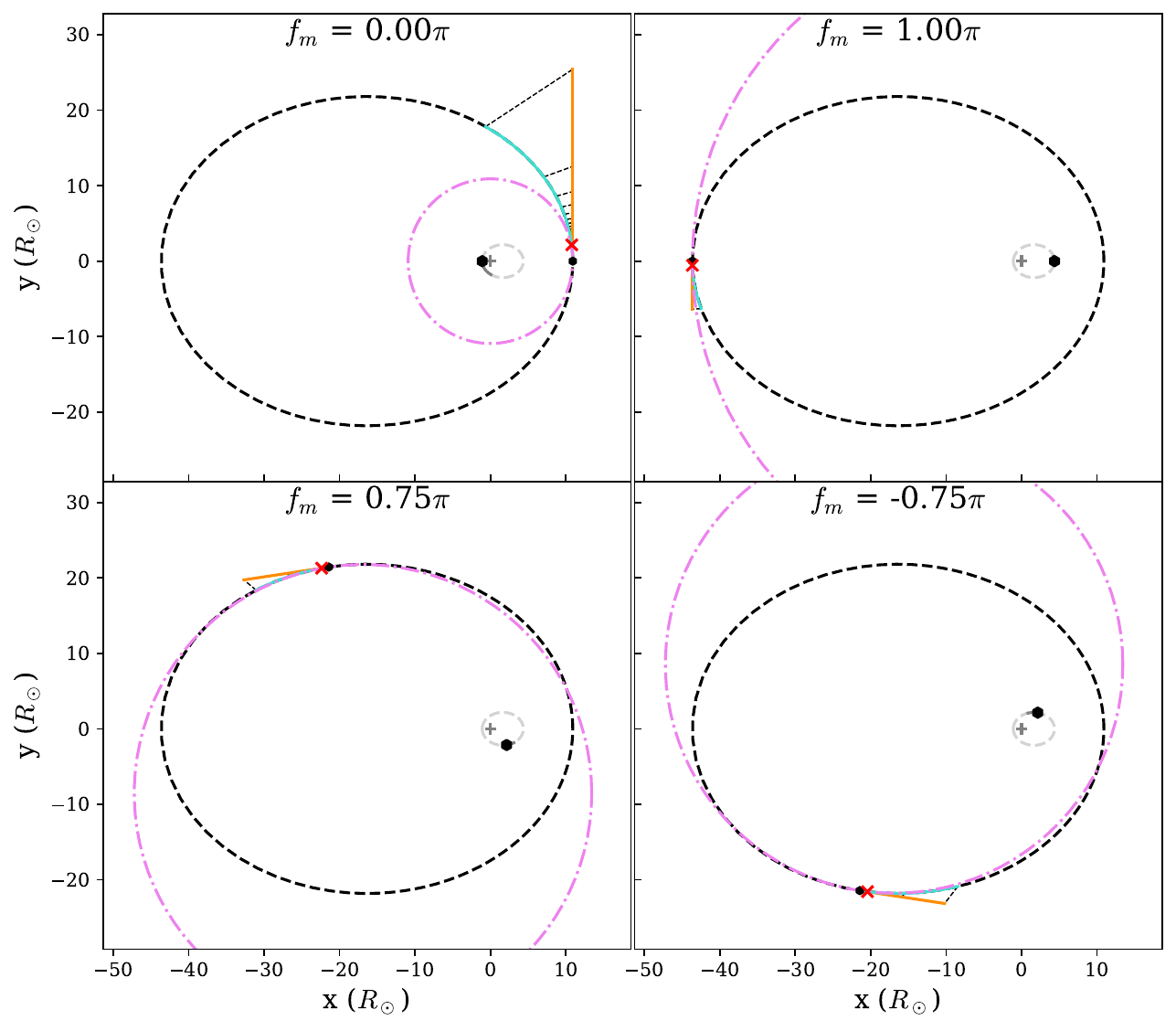}
    \includegraphics[width=.8\textwidth]{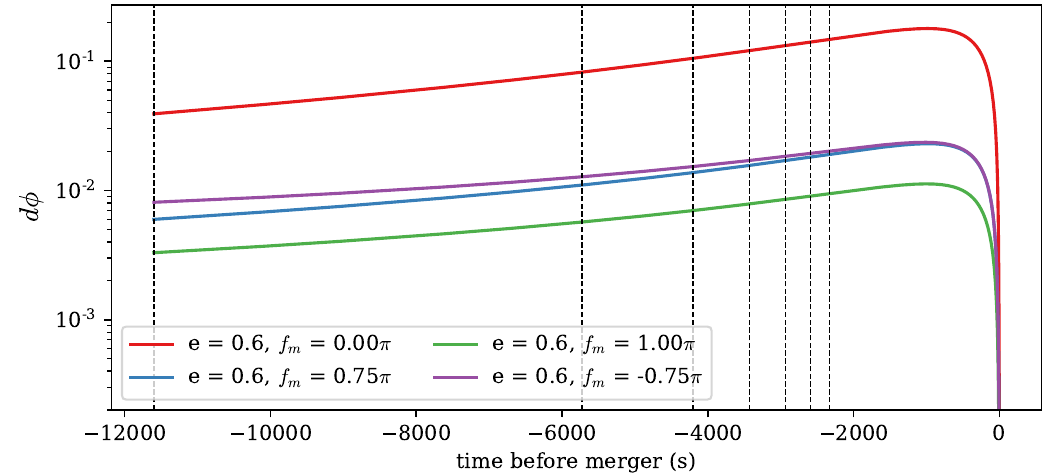}
    \caption{{\bf GW phase shifts from eccentric orbits.} \textit{Top 4 panels:} Trajectories of four different GW phase shift-inducing scenarios, each with different $f_m$. The eccentricity of the outer orbit is set to $e_{\rm out} = 0.6$. The turquoise curve is the true trajectory of the binary (merging at the smaller black dot at the corresponding $f_m$) on an eccentric orbit around BH3 (the larger black dot). The reference trajectory in isolation is depicted by the straight orange line. The dashed straight lines represent the distance $l(t)$ between the two trajectories which is used to calculate the time delay according to $\Delta t(t) = l(t)/c$. The red cross corresponds to the place on the orbit where $d\phi_\mathrm{max}$ occurs, which is quite close to merger in all cases. We also show in pink the circle with radius equal to $\mu r_m$, which we use for our circular approximation. The black and grey dashed ellipses depict the Keplerian orbits of the binary COM and BH3, respectively. \textit{Bottom panel:} the phase shift $d\phi$ as a function of time for each scenario. The black dashed lines in this panel represent the temporal location of the same lines in the trajectory plots.}
    \label{fig:varying_fm_eccp6}
\end{figure*}
\begin{figure*}
    \centering
    \includegraphics[width=.8\textwidth]{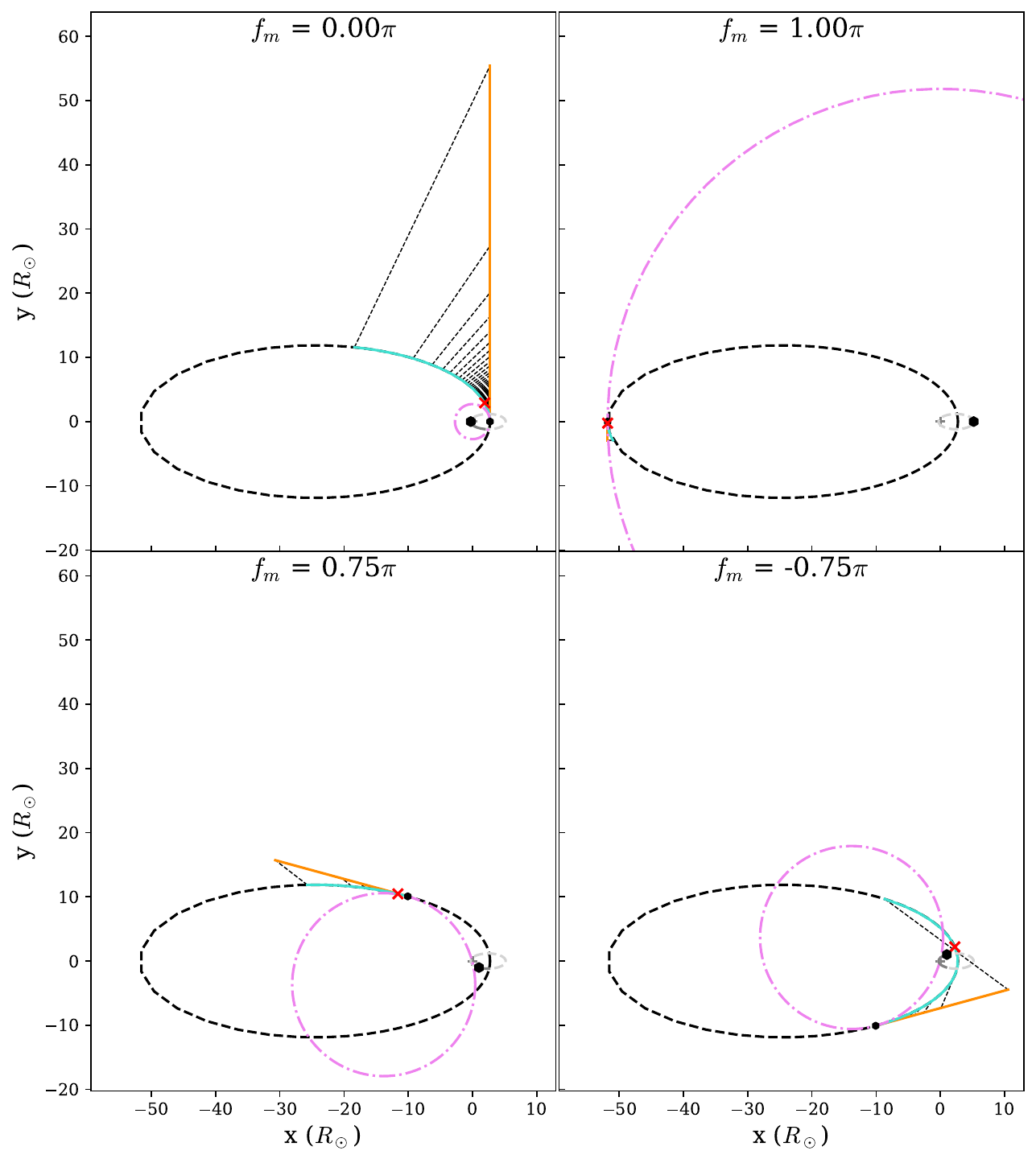}
    \includegraphics[width=.8\textwidth]{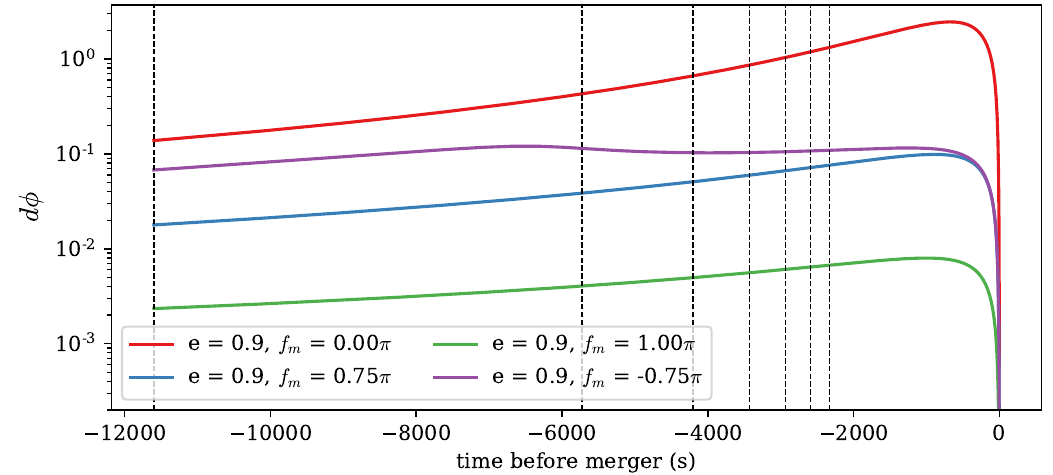}
    \caption{{\bf GW phase shifts from eccentric orbits.} Similar to Fig. \ref{fig:varying_fm_eccp6}, but with the eccentricity of the outer orbit set to $e_{\rm out} = 0.9$.}
    \label{fig:varying_fm_eccp9}
\end{figure*}

\subsubsection{Fast Linear Approximation}\label{sec:Fast Estimator Phaseshift}

The fastest estimator for the maximum phase shift $d{\phi}$ can be calculated using Eq. \ref{eq: dphi_gen_a} with
\begin{equation}
|{\bf a}| = \frac{Gm_3}{r_{m}^2}
\end{equation}
from which it directly follows,
\begin{align}
\begin{split}
    d\phi(e) &\approx \frac{1}{2} \frac{G^{3/2}}{c} \frac{m_3 m_{12}^{1/2}}{r_m^2} \frac{t(e)^2}{a(e)^{3/2}} \\
    &= \frac{1}{2} \frac{G^{3/2}}{c} m_3 m_{12}^{1/2} \\
    &\qquad \times \left(\frac{1+e_\text{out}\cos f_m}{a_\text{out}(1-e_\text{out}^2)}\right)^2 \frac{t(e)^2}{a(e)^{3/2}},
    \label{eq:dphi_small}
\end{split}
\end{align}
where in the last equality we have made the relation to the orbital elements describing the eccentric outer
trajectory of the BBH relative to the single BH. Note that this equation also can be deduced from Eq. \ref{eq:dt_circ},
by taking the small angular limit ${\theta} << 1$. One general result that follows from this is that the maximum phase shift,
also in this outer orbit eccentric case, scales $\propto 1/{r_m}^2$, i.e. the closer to pericentre the merger happens,
the higher the phase shift. This is one of the main reasons and motivations to study the outer eccentric case as we do here.

While one can substitute $a$, $e$, and $t$ in the above equations, it is not easy to read off the maximum of $d\phi$. However,
by combining this linear approximation and using Peters64 for both the inspiral time $t$ and for $a(e)$ it becomes possible
\citep{samsing_gravitational_2024}, even here in the more general case of an outer eccentric orbit.
Following this approach, we now start by substituting $t$ in Eq. \ref{eq:dphi_small} with the approximative estimator given by Peters64, 
\begin{equation}
t_{c} = \frac{3/85}{c^5/G^{3}} \frac{a^{4}}{m_{1}m_{2}m_{12}} \times (1-e^{2})^{7/2}.
\label{eq:tmP64}
\end{equation}
The reason why we use this expression here instead of the fit Eq. \ref{eq:tc} is that this equation, despite being slightly less accurate,
allows for direct estimation of the binary eccentricity at which the maximum phase shift occurs.
This then results in an analytical expression for the maximum phase shift and its relevant scalings of the problem,
which is extremely valuable for studies with e.g. N-body simulations containing millions of few-body interactions and mergers.

For $a(e)$ we use Eq. \ref{eq:a_e} and Eq. \ref{eq:ge} that in the limit where the BBH is assembled with a high initial eccentricity, $e_0$,
can be written as \citep[see][]{samsing_gravitational_2024},
\begin{equation}
a(e) \approx \frac{2r_0e^{12/19}}{(1-e^2)}\frac{h(e)}{h(1)},\ (e_0 \approx 1),
\label{eq:ae_e1lim}
\end{equation}
where $r_0$ is the initial pericentre distance at assembly of the inspiralling BBH and 
\begin{align}
\begin{split}
    h(e) &= \bigg(1 + \frac{121}{304}e^2\bigg)^{870/2299} \\
    &= g(e) \frac{1 - e^2}{e^{12/19}}.
\end{split}
\end{align}
The input $r_0$ is well defined and easy to estimate, in contrast to $a_0$ and $e_0$.
Combining these relations, we reach the following relation,
\begin{align}
	d{\phi}(e) & \approx \frac{288\sqrt{2}}{85^{2}h(1)^{13/2}} \frac{c^{9}}{G^{9/2}} \times \frac{m_3}{r_m^2}\frac{r_0^{13/2}}{m_1^{2}m_2^{2}m_{12}^{3/2}} \nonumber\\
	 	    &  \times e^{78/19}(1-e^2)^{1/2}h(e)^{13/2}.
    \label{eq:dphi_e_e1lim}
\end{align}
The maximum value of $d{\phi}$ can now easily be found by maximizing the function
\begin{equation}
F(e) = e^{78/19}(1-e^2)^{1/2}h(e)^{13/2},
\label{eq:FunctionFe}
\end{equation}
from which one finds
\begin{equation}
e_m = \sqrt{{2\left(\sqrt{391681} - 115 \right)}/{1213}} \approx 0.92,
\label{eq:e_maxphi}
\end{equation}
where $e_m$ denotes the value that maximizes $F(e)$ and thereby $d{\phi}$. 
From this we conclude that the maximum possible value for $d{\phi}$ for a dynamically assembled BBH that mergers at
a distance $r_m$ from a perturber can be put in closed form as,
\begin{align}
	d{\phi}_\mathrm{max} & \approx \frac{288\sqrt{2}}{85^{2}h(1)^{13/2}} \frac{c^{9}}{G^{9/2}} \times \frac{m_3}{r_m^2}\frac{r_0^{13/2}}{m_1^{2}m_2^{2}m_{12}^{3/2}} \nonumber\\
	 	    &  \times e_m^{78/19}(1-e_m^2)^{1/2}h(e_m)^{13/2},
       \label{eq:max_dphi}
\end{align}
with $e_m \approx 0.92$.
One should note here that this maximum value might not be in the observable band. As was proven in \cite{samsing_gravitational_2024}, the initial
BBH peak frequency $f_0(r_0)$ needs to be near the observable band for the phase shift effects to be large enough to be observable, i.e. for
these calculations to be relevant for e.g. LIGO, $f_0$ at assembly has to be $ \sim 10$ Hz or above. It can be lower, but then the phase shift
will have decreased significantly before reaching the observable band.
For example, in the limit where the BBH is assumed circular, $d{\phi} \propto f^{-13/3}$, i.e. it rapidly drops as the frequency increases.

\subsection{Limitations and Conclusions}\label{sec:conclude app}

Both the linear and the circular approximations from above are based on assumptions of constant velocity and acceleration during
the time over which the phase shift is observed, which are generally justified. However,
in some cases, especially relevant for the LISA mission, the observational time
window can be comparable or even longer than the outer orbital time, which leads to periodic features from the time-evolving
Doppler shift that will naturally enter the GW waveform and phase shift. In our considered case, where
dynamically assembled binaries are often evolving along highly eccentric outer orbits, other effects can occur on much smaller timescales, as
the acceleration will vary greatly over the orbit from pericentre to apocentre. Most notably, if the BBH happens to pass near
pericentre at the time of observation, one expects large non-linear dephasing effects on a timescale set by the time of pericentre passage, which
can be orders-of-magnitude smaller than the outer orbital time. Such effects are not captured by our approximations from this section,
but below we will study such encounters in much more detail.

\section{Phase Shift Characteristics}\label{sec:phase_shift}

Here, we explore the unique features arising in the GW phase shift
from changing the outer orbit eccentricity $e_\mathrm{out}$, and the angular position at which merger takes place, $f_m$,
for chirping and merging eccentric BBHs. For the illustrative examples below, unless otherwise stated, we consider cases for which $m_1=m_2=5 M_\odot$, $m_3 = 100 M_\odot$, $a_0 = 1.3 R_\odot$, $e_0 = 0.999$, and $a_\mathrm{out} = 30 R_\odot$. These values for the inner BBH correspond to a GW burst source with peak frequency $f_{p,GW} \approx 32$Hz.

\subsection{Imprints of Orbital Eccentricity}\label{sec:mod_ecc}

In Fig. \ref{fig:varying_fm_eccp6} we show the BBH evolution along outer orbits with a representative moderate eccentricity of $0.6$, as well as with reference binary
tracks, for 4 different values of $f_m$. The corresponding GW phase shift, $d\phi$, as a function of time to merger,
is depicted in the lowest panel of the figure. For all these computations we follow our procedure outlined in Sec. \ref{sec:numerical}.

The general shapes of the curves in Fig. \ref{fig:varying_fm_eccp6} are reminiscent of the circular case \citep{samsing_gravitational_2024},
i.e., the peak that is present in the phase shift of circular outer orbits also persists here in the eccentric case.
However, the evolution and peak value of $d\phi$ clearly vary with $f_m$, as the value of $f_m$ maps to $r_m$ that determines the
magnitude of the phase shift, $d\phi_\mathrm{max} \propto r_m^{-2}$, which follows from Eq. \ref{eq:max_dphi}.
This also implies that when the outer orbit is eccentric, the maximum value $d\phi_\mathrm{max}$ will change by a factor
set by the pericentre and apocentre distances, such that
\begin{align}
	\frac{d{\phi}_\mathrm{peri}}{d{\phi}_\mathrm{apo}} \approx \frac{(1+e)^2}{(1-e)^2},
    \label{eq:dphi_diff}
\end{align}
and correspondingly by a factor of $1/(1-e)^2$ compared to if the BBH was circular with radius $a_\mathrm{out}$.

Note how, for the cases $f_m = -0.75\pi$ and $f_m = 0.75\pi$, the distance at merger between the binary and the third object is the same. Therefore these cases have an almost identical peak phase shift. However, their phase shifts diverge at earlier times because of the differing ellipse curvatures along the orbit.

\begin{figure}
    \centering
    \includegraphics[width=.47\textwidth]{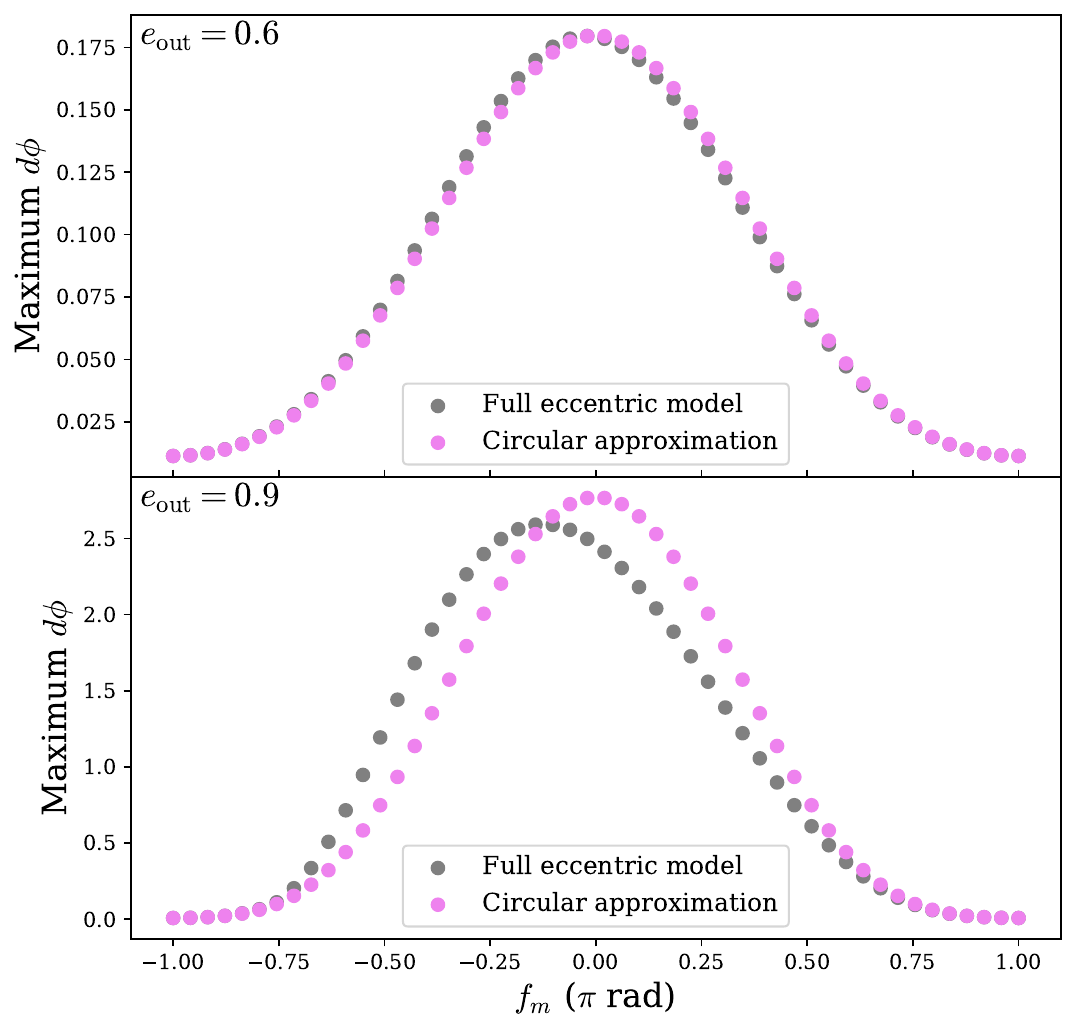}
    \caption{{\bf Maximum GW phase shift.} \textit{Top}: In grey, we show the dependence of $d\phi_\mathrm{max}$ on the true anomaly at merger $f_m$. The pink curve represents our circular approximation (Eq. \ref{eq:dphi_circ}). The outer eccentricity is 0.6. \textit{Bottom}: the same relation, this time with $e_\mathrm{out}=0.9$.}
    \label{fig:phim_fm}
\end{figure}

In summary, the leading order effect from introducing a small to moderate eccentricity of the outer orbit, is a change in the maximum phase
shift $d\phi_\mathrm{max}$. Generally, as the distance $r$ between the binary and third object at the point of merger increases, the peak value
decreases as $\propto 1/r^2$. Note here that there will be further dynamically constraints on the BBH if the outer orbit eccentricity
increases. We will touch upon this in Sec. \ref{sec:tides}.

\subsection{Strong Effects and Secondary Peaks}\label{sec:high_ecc}

Having studied features of the moderately eccentric case, we now move to the case where the outer orbit has a significant eccentricity,
which we show will give rise to additional unique imprints on the GW phase shift. For this, we start by considering Fig. \ref{fig:varying_fm_eccp9},
which shows the trajectories of the BBH (with respect to the isolated case) at four different value of $f_m$, with an outer eccentricity of 0.9.

We focus here on the new qualitative features that go beyond the expectations from the $1/r_m^2$ scaling. For highly eccentric outer orbits, the maximum phase shift is achieved for binaries that merge shortly after their pericentre passage. As an example, a binary merging at $f_m \sim -0.2\pi$ passes the high curvature region at pericentre, changing direction rapidly. This change in direction, combined with a high orbital velocity,
causes higher possible phase shifts as opposed to a binary simply merging at pericentre ($f_m = 0$). In the regions $-0.75\pi < f_m < -0.25\pi$ and $0.25\pi < f_m < 0.75\pi$, the circular approximation respectively under- and overestimates the maximum phase shift due to the extra ac- and deceleration of the binary from the strong curvature. This is futher outlined in Fig. \ref{fig:phim_fm}.

\begin{figure}
    \centering
    \includegraphics[width=.47\textwidth]{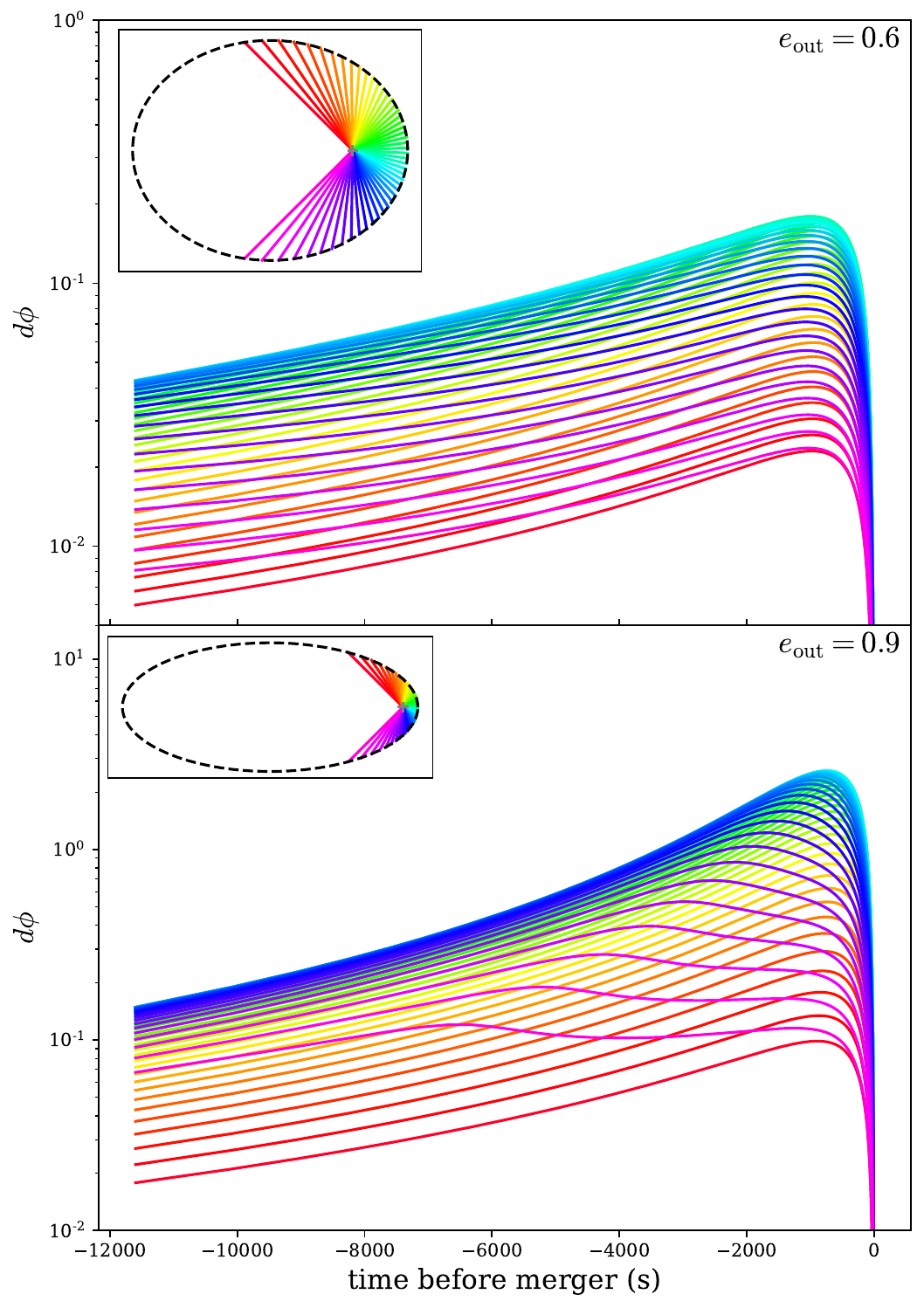}
    \caption{{\bf GW phase shift, eccentricity, and double peaks.} The phase shift $d\phi$ as a function of time, for a range of $f_m$ close to pericentre ($-0.75\pi \leq f_m \leq 0.75\pi$). The ellipse in the top left corner illustrates the outer eccentric orbit. Every coloured line represent a certain $f_m$, where the location of merger is the point where that line intersects with the ellipse. The colour of each line in the main plot corresponds to the colour of the lines within the ellipse. {\textit{Top:}} We show the results for $e_\mathrm{out} = 0.6$. {\textit{Bottom:}} Results for $e_\mathrm{out} = 0.9$.}
    \label{fig:dphi_t_shaded}
\end{figure}

\begin{figure}
    \centering
    \includegraphics[width=0.45\textwidth]{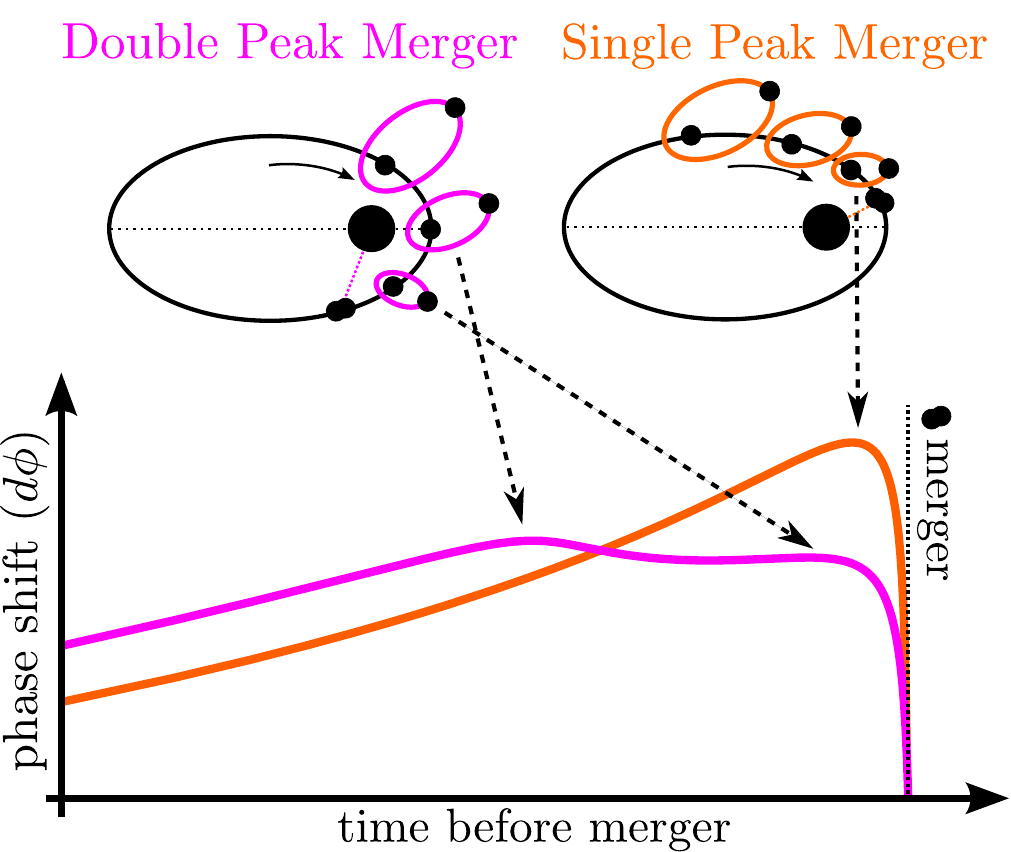}
    \caption{{\bf Formation of double peak mergers.} Schematic visualisation of the secondary peak. In orange, we depict the trajectory of a binary merging before pericentre, giving rise to the typical single-peak phase shift. The trajectory of a binary merging after pericentre is shown in pink, where the pericentre passage gives rise to a second peak.}
    \label{fig:2peaks}
\end{figure}

More characteristics unique to highly eccentric outer orbits are visible in Fig. \ref{fig:varying_fm_eccp9}.
The case $f_m=-0.75\pi$ has two peaks: a main peak that occurs at around the same time as for the other curves, and a secondary peak
appearing at earlier times. In Fig. \ref{fig:2peaks} we schematically depict this scenario, where in orange we show the trajectory
of a binary that produces a single-peak phase shift, and in pink a case where the merger happens right after peri-center passage,
which gives rise to the earlier second peak. There is an $f_m$ sweetspot
on the outer orbit where the secondary peak may actually be higher than the main one. For this to happen, $f_m$ needs to be far enough
from pericentre that the pericentre passage happens \textit{before} the main peak, but close enough to pericentre that the distance
is not to large. In our considered setup, the sweet spot is at $f_m \sim -0.2 \pi$. We stress that this is an important feature
of a system like this, as it is a direct mapping of the environment in which the BBH merges.
This only occurs when $e_\mathrm{out}$ is high and only occurs in a specific part of the outer orbit so a possible
observation of this would undoubtedly pin down properties of this BBH formation channel.

In Fig. \ref{fig:dphi_t_shaded} we show in more detail how the GW phase shift curves vary around pericentre with changing $f_m$, where each colored line
corresponds to a different value of $f_m$, as illustrated in the upper left insert figure for each panel ($f_m$ for a
given coloured line is where the line ends at the outer orbit shown with a black dashed line). In the top panel,
we see that curves with the highest peak in the $e_\mathrm{out}=0.6$ scenario are green/light blue, which correspond to mergers near pericentre.
The colour of the highest peak at $e_\mathrm{out}=0.9$ is a deeper blue, corresponding to mergers just after pericentre (at $f_m\sim-0.2\pi$).
The bottom panel visualises how the secondary peak evolves. As we move from pink into blue, the merger happens increasingly close to
pericentre, so the secondary peak slowly melts together with the main peak.

\begin{figure}
    \centering
    \includegraphics[width=.5\textwidth]{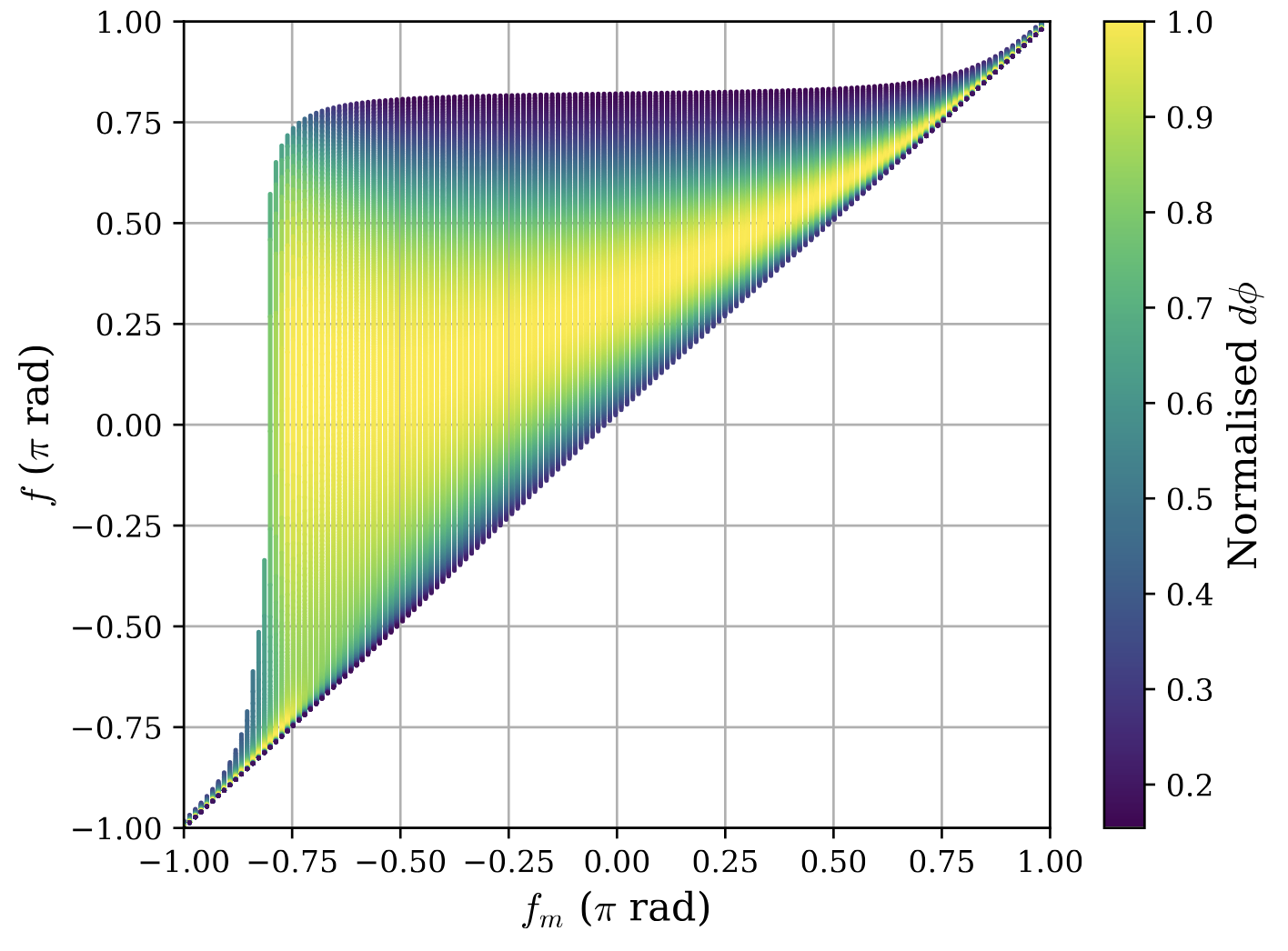}
    \caption{{\bf Orbital phase and GW phase shift.} Heat map of the the phase shift, as a function of the orbital phase $f$, for different realisations of $f_m$. Each vertical line represents the trajectory of an inspiralling binary, forming at some true anomaly $f_f$ (top), and merging at $f_m$ (bottom). We normalise $d\phi$ for each trajectory, such that $d\phi_\mathrm{max}=1$ for every $f_m$. We employ an outer eccentricity of 0.9.}
    \label{fig:dphi_t_heatmap}
\end{figure}

Fig. \ref{fig:dphi_t_heatmap} visualises the shape of $d\phi$, including the changing location of the maximum, for a full
range of $f_m$ at a high outer eccentricity. Each vertical line in this plot represents the GW phase shift path of a binary merging at a
specific $f_m$, where the binary evolves as a function of true anomaly $f$ from the top and down.
The colour represents the (normalised) phase shift at each $f$ in the orbit. It is clearly seen how the evolving GW phase shift
depends on how and where the BBH merges on the outer eccentric orbit. In particular, between $f_m\sim-0.75\pi$ and $f_m \sim -0.50\pi$
the presence of the secondary peak is clearly visible.

To sum up, outer orbits with high eccentricities give rise to extra features and effects that are not present otherwise.
Firstly, an increased eccentricity results in a much closer encounter at pericentre which leads to significantly larger phase shifts.
Additionally, strong curvature effects around pericentre leave their imprint on the shape of the phase shift:
$(i)$ The largest possible phase shift does not occur when the binary merges at pericentre, but slightly
beyond (at $f_m\sim-0.2\pi$ in our example) due to the extra curvature effects due to passing pericentre, and $(ii)$ a
secondary peak shows up when the binary passes pericentre before it reaches the main peak (for $f_m\sim-0.75\pi$ in our setup).
Although these effects are lost in our linear approximations from Sec. \ref{sec:approx}, we find that it still quite
accurately predicts the magnitude of the peak.

\subsection{Generalised Behaviour}

\begin{figure}
    \centering
    \includegraphics[width=.5\textwidth]{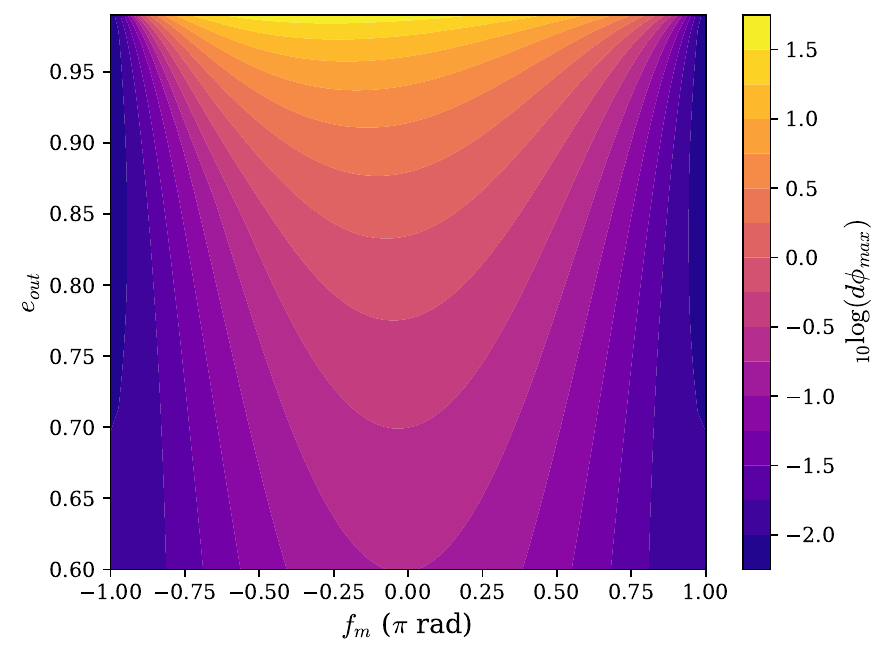}
    \includegraphics[width=.48\textwidth]{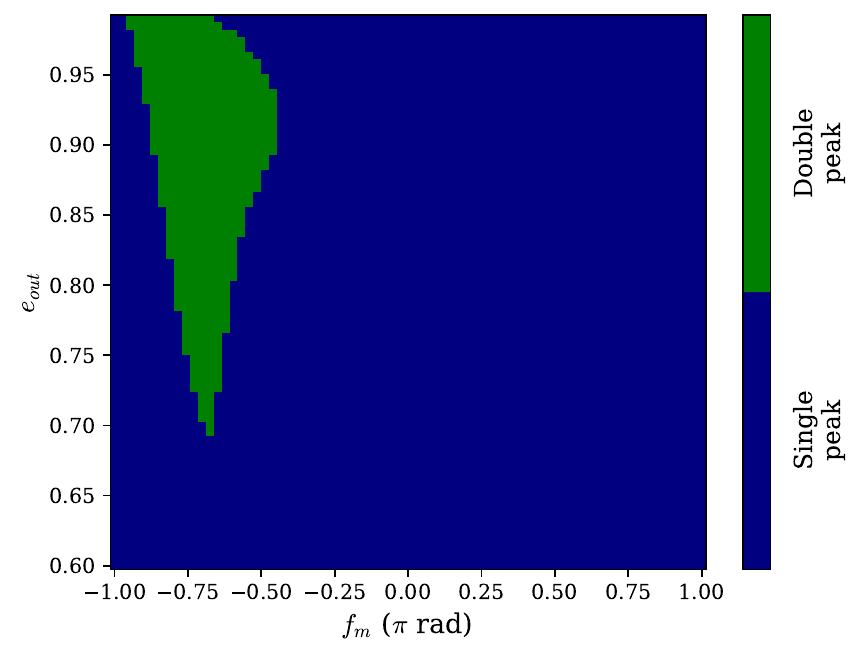}
    \caption{{\textit{Top:}} Maximum phase shift $d\phi_\mathrm{max}$ as a function of $f_m$ and $e_\mathrm{out}$ for the parameters listed in Sec. \ref{sec:phaseshift}. {\textit{Bottom:}} Phase space area in $f_m$ and $e_\mathrm{out}$ that presents a double-peaked (green) or a single-peaked (blue) dephasing curve.}
    \label{fig:phim_contour}
\end{figure}

We generalise our results more by showing the transition between moderate and high eccentricities and when strong curvature effects begin to occur.
In the top panel of Fig. \ref{fig:phim_contour} we show $d\phi_\mathrm{max}$ as a
function of $f_m$ for a range of $e_\mathrm{out}$ (from $0.6$ to $0.99$) for our considered system.
Each horizontal slice of this plot can be viewed as a curve
like Fig. \ref{fig:phim_fm}. It becomes clear that eccentricity effects are already visible as low as $e_\mathrm{out}=0.6$. There,
the location of the maximum phase is already shifted to the left of periastron. Between
$e_\mathrm{out}=0.6$ and $e_\mathrm{out}=0.95$, the maximum possible phase shift steadily increases by about 3 orders-of-magnitude.

It is further informative to show the emergence of the secondary peak as we increase the outer eccentricity, and in what range of $f_m$
the peak appears. We portray this in the bottom panel of Fig. \ref{fig:phim_contour}, where the $x$-axis shows $f_m$ and the $y$-axis $e_\mathrm{out}$.
We define the presence of a double peak by a sign change in the second time derivative of the phase shift. The green area shows the region
of $f_m$ and $e_\mathrm{out}$ for which there is a secondary peak. The double peak does not occur for the lowest eccentricities,
as the curvature effects of passing pericentre are not large enough. With this setup, the secondary peak appears for $e_\mathrm{out}$
as low as $\sim0.7$. As $e_\mathrm{out}$ increases the green region widens. On the left, it asymptotes towards apocentre.
On the right, the width increases faster due to the strong curvature effects. Above $e_\mathrm{out}\sim0.9$, the width decreases again, as the binary moves extremely fast close to pericentre. For an eccentricity of $\sim0.95$,
the time between pericentre passage and a merger happening at $f_m$ of $\sim-0.5\pi$ is so short that the two peaks become indistinguishable.

In summary, the features of the phase shift that are typical of triples with a high outer eccentricity can show up in the phase
shift at moderate to moderately high eccentricity.
In terms of detections this is a positive result as it indicates towards the possibility of distinguishing circular outer orbits from
their eccentric counterparts not just in the most extremely eccentric cases.

\section{Astrophysically Relevant Scenarios}\label{sec:astro}

Here we showcase two examples of astrophysical scenarios that may give rise to potentially detectable phase shifts with interesting features.
We first investigate in more detail the case of Fig. \ref{fig:scatteringEX1}, i.e. a triple scattering in e.g. a stellar cluster \citep[see e.g.][]{Samsing18, 2006ApJ...640..156G, 2014ApJ...784...71S, 2017ApJ...840L..14S}. Then, we discuss the influence of tidal forces from the perturber onto the amplitude of the phase shift.

\subsection{Three-body Scatterings}\label{sec:scattering}

\begin{figure}
    \centering
    \includegraphics[width=.48\textwidth]{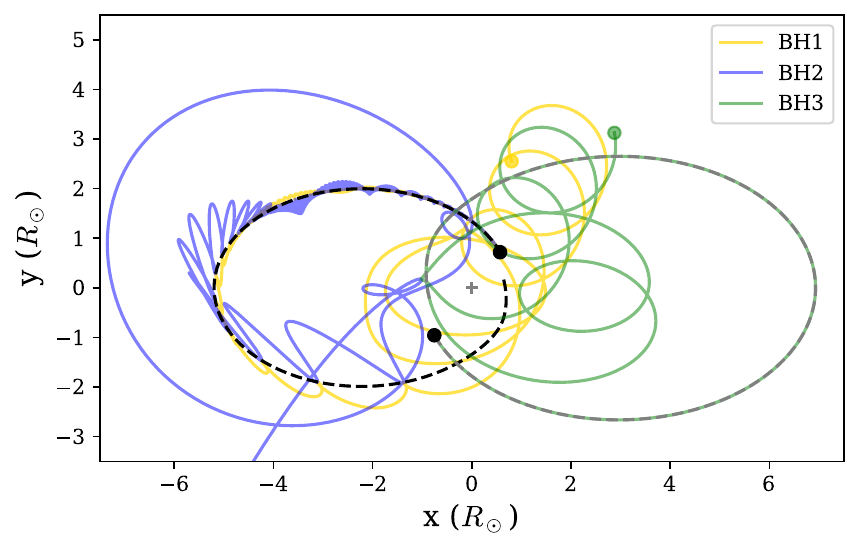}
    \includegraphics[width=.48\textwidth]{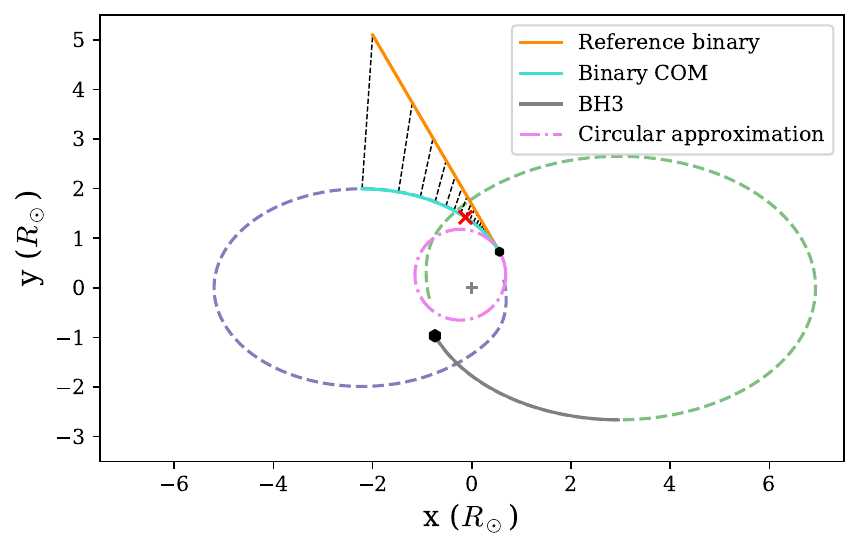}
    \caption{{\bf Eccentric orbits in 3-body interactions.} \textit{Top:} The trajectories of the scattering from Fig. \ref{fig:scatteringEX1} for illustration. \textit{Bottom:} the binary and reference trajectories including Rømer delay for this particular scattering, including the orbit for our circular approximation. The shaded dashed lines show the final orbits of the binary (blue) and third object (green) in the N-body simulation. The location of the maximum phase shift is depicted by the red cross. All trajectories are shown in the COM frame of the triple.}
    \label{fig:Nbody_scattering}
\end{figure}

Chaotic scatterings are ideal environments for producing systems with potentially detectable GW phase shifts.
In the top panel of Fig. \ref{fig:Nbody_scattering}, we show once again the trajectories of a chaotic triple producing a
highly eccentric merger. From the positions and velocities at merger we can extract the orbital elements of the triple
that are needed as input for our model to compute the phase shift, except for $a_0$ and $e_0$ which we extract at some reference
time before merger. In the bottom panel of Fig. \ref{fig:Nbody_scattering} we plot the binary and reference trajectory
(blue and orange) on top of the data from the N-body simulation (grey shaded). 

\begin{figure}
    \centering
    \includegraphics[width=.48\textwidth]{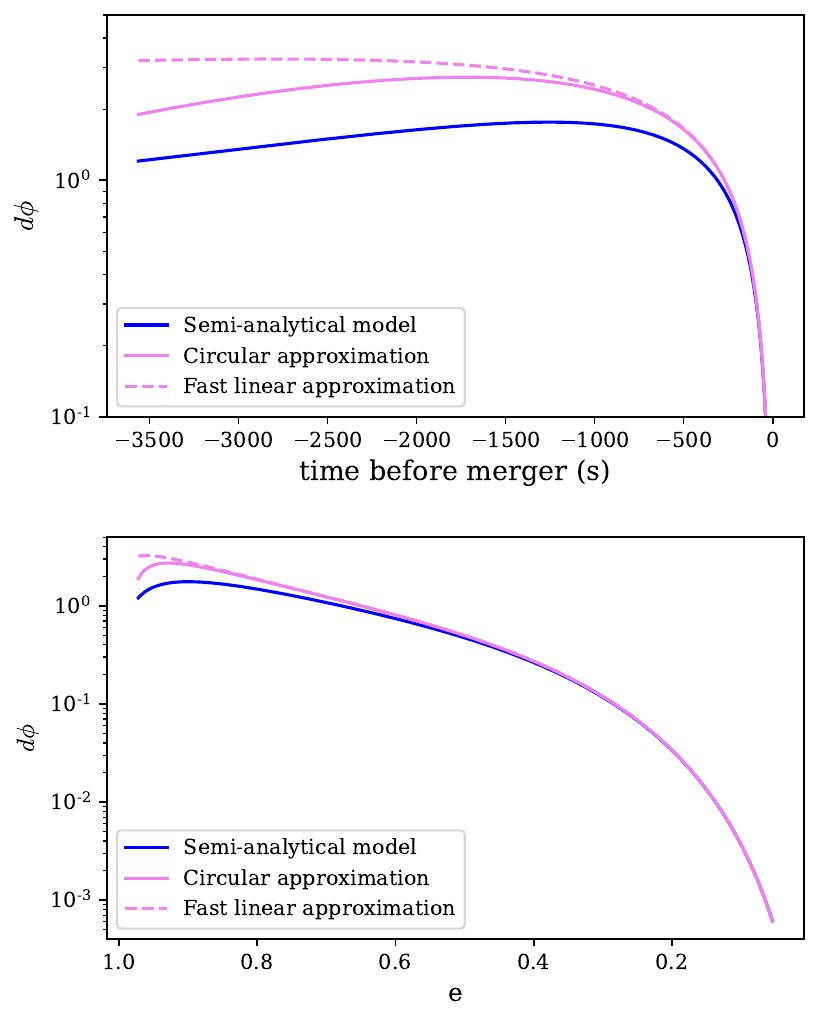}
    \caption{{\bf 3-body GW phase shifts and models.} GW phase shift as a function of time (\textit{top}) and orbital eccentricity (\textit{bottom}) for the case study depicted in Fig. \ref{fig:Nbody_scattering}. In blue, we show the result from our semi-analytical model, and in pink our approximations: the solid line depicts the circular approximation (Eq. \ref{eq:dphi_circ}), and the dashed line shows the fast linear approximation (Eq. \ref{eq:dphi_e_e1lim}).}
    \label{fig:dphi_Nbody}
\end{figure}

The resulting phase shift $d\phi$
of this scattering is depicted in Fig. \ref{fig:dphi_Nbody}, as a function of time (top) and orbital eccentricity (bottom).
We observe that both versions of our approximation slightly overestimate the magnitude of the (maximum) phase shift, as compared to
the prediction from our full semi-analytical model. The reason for this is simply that for this setup, and in this part of the outer orbit, the
acceleration when the phase shift is at its maximum (red cross) has changed significantly compared to the value at merger.
However, despite these known limitation of our analytical approximations, we do find that the estimates for $d\phi_\mathrm{max}$ from
both the fast linear approximation (Eq. \ref{eq:max_dphi}) and the circular approximation (Eq. \ref{eq:dphi_circ}) are only about a factor of $1-3$
off compared to our more accurate semi-analytical model. This greatly motivates to explore the possibilities of using such approximations
for future large-data projects involving determining GW phase shifts for millions of PN few-body scatterings across cosmic time for e.g.
different cluster environments. Lastly, we see that our estimates here are fully consistent with the numerically estimated phase shift for this
scattering shown in \cite{samsing_gravitational_2024}. Especially our semi-analytical model correctly captures the
slope of $d\phi$ on both sides of the peak, whereas the circular approximation that was over-plotted in \cite{samsing_gravitational_2024} deviates from the truth at earlier times.


\subsection{Influence of Tides}\label{sec:tides}

BBHs assembled during chaotic scatterings are often undergoing their first part of their inspiral while being subject to tidal influence from the nearby perturber. This effect was studied for a few numerical cases in \cite{samsing_gravitational_2024}. Here we perform a more controlled study of this, by investigating how the tidal influence on the eccentricity of the inspiralling BBH will propagate to changes in the expected later GW phase shift. As described below, for this we consider a BBH (BH1,BH2) that is assembled near the Hill sphere with respect to a third object (BH3).

For our initial setup, we consider two BHs that become bound near BH3 after radiating an amount $\Delta E_{\text{GW}}$ of GW
radiation during their first pericentre passage.
We can describe this as \citep[see][]{peters_gravitational_1964, hansen_post-newtonian_1972, samsing_dissipative_2018}:
\begin{align}
    \Delta E_{\text{GW}} = \frac{85\pi}{12\sqrt{2}} \frac{G^{7/2}}{c^5} \frac{m_1^2 m_2^2 m_{12}^{1/2}}{r_p^{7/2}},
    \label{eq:deltae_gw}
\end{align}
Here, $r_p = a(1-e)$ is the distance at pericentre, and $m_{12} = m_1+m_2$. Under the assumption that the two BHs
will end up near the Hill sphere relative to BH3 after their first pericentre passage, they will have an orbital energy given by,
\begin{align}
\begin{split}
    \Delta E_{\text{orb}} &= \frac{Gm_1m_2}{2a} \approx \frac{Gm_1m_2}{R_H}.
\end{split}
    \label{eq:deltae_orb}
\end{align}
The definition of the Hill radius of BH1, on which we place BH2, is
\begin{align}
\begin{split}
R_H &= r(f)\left(\frac{m_1 + m_2}{3m_3}\right)^{1/3} \\
&= \frac{a_{\text{out}}(1-e^2_{\text{out}})}{1+e_{\text{out}}\cos f}\left(\frac{m_1+m_2}{3m_3}\right)^{1/3}.
\end{split}
\end{align}
By combining Eqs. \ref{eq:deltae_gw} and \ref{eq:deltae_orb} we now have an estimate for the initial eccentricity and semi-major axis
in this GW capture scenario, which is a toy description of what do realistically happen in triple scatterings,
\begin{align}
    e_0 \approx 1 - 2 \left(\frac{85\pi}{12\sqrt{2}}\right)^{2/7} \frac{G^{5/7}}{c^{10/7}} m_1^{2/7} m_2^{2/7}m_{12}^{1/7} \frac{1}{R_H^{5/7}},
    \label{eq:e0}
\end{align}
and
\begin{align}
    a_0 \approx \frac{R_H}{2}.
    \label{eq:a0}
\end{align}

\begin{figure*}[t]
    \centering
    \includegraphics[width=1.0\textwidth]{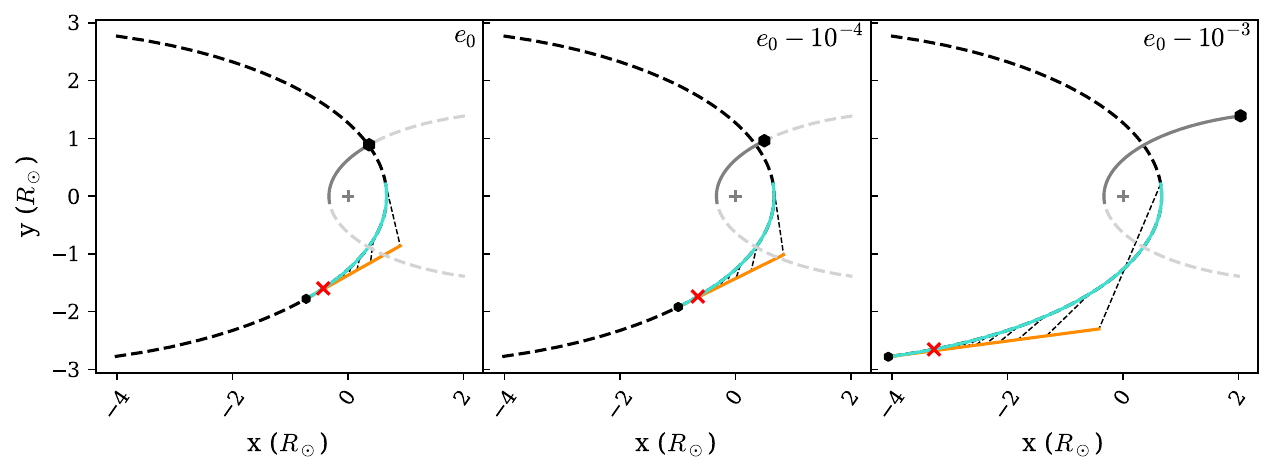}
    \caption{{\bf Orbital changes from tides.} Trajectories of the binary COM (turquoise), reference binary (orange), and third BH (grey), for 2 different values subtracted from $e_0$ ($10^{-4}$, and $10^{-3}$). The choices for the other parameters are described in Sec. \ref{sec:tides}. }
    \label{fig:traj_tides}
\end{figure*}

\begin{figure}
    \centering
    \includegraphics[width=.48\textwidth]{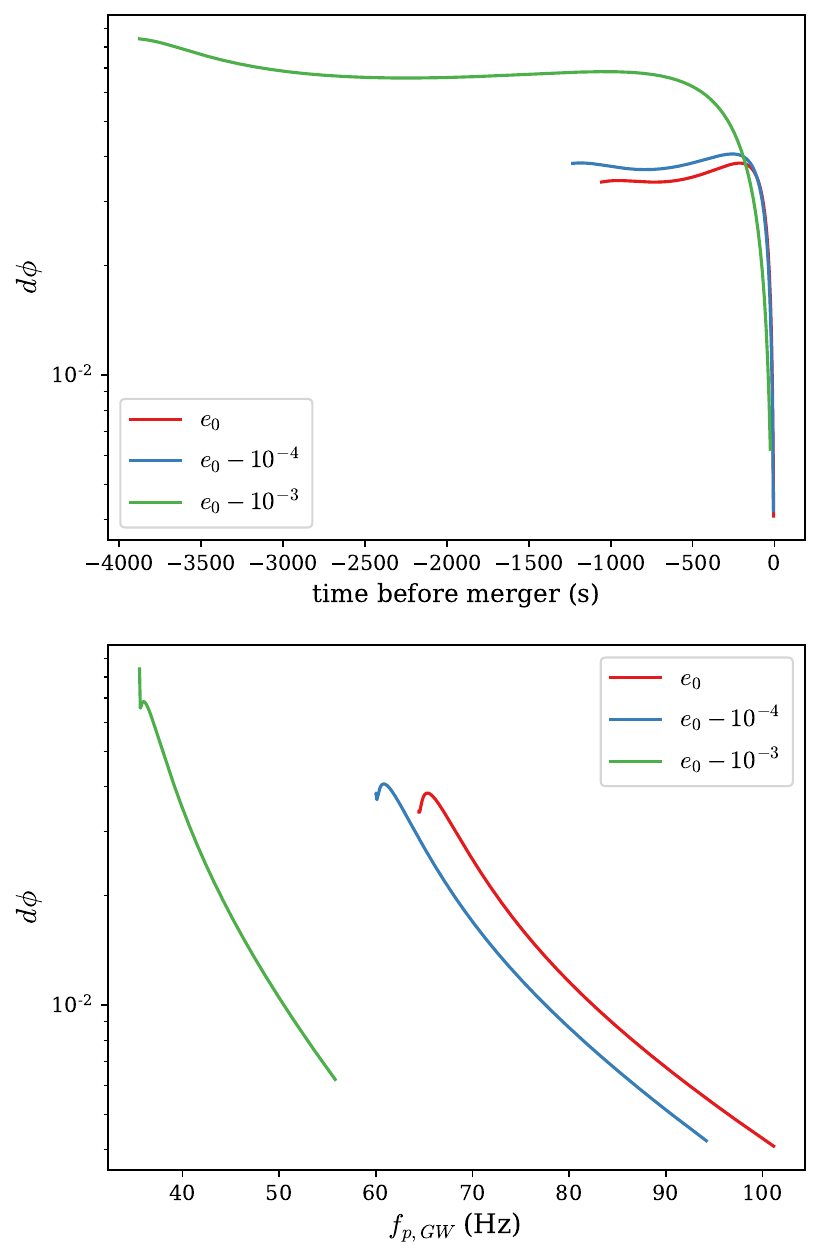}
    \caption{{\bf Tidal influence on GW phase shift evolutions.} Phase shift as a function of time (\textit{top}) and peak GW frequency (\textit{bottom}) for three different initial binary eccentricities. We start with $e_0$ according to Eq. \ref{eq:e0}, and systematically subtract a small value to alter the phase shift evolution. We use $m_1=m_2=5M_\odot$, $m_3=20M_\odot$, $e_\mathrm{out}=0.9$ and $a_\mathrm{out}=10R_\odot$. We observe that each of these trajectories shows the secondary peak in the phase shift that was discussed in the above.}
    \label{fig:dphi_tide}
\end{figure}

Upon formation of the inspiralling BBH, when the binary is still relatively wide, tidal forces from the third object can alter the
initial semi-major axis and eccentricity \citep[see also][]{samsing_gravitational_2024}. Exactly how and to what extent the binary is
influenced by the tides is chaotic and is dependent on several factors, such as the orientation and distance with respect to the third object.
Any change to the initial conditions may have a large effect on the inspiral time and thereby evolution of the phase
shift. To illustrate this, here we investigate how small changes in the initial binary eccentricity $e_0$ may affect the phase shift evolution.
We create a binary that forms on the Hill radius with eccentricity and semi-major axis according to Eqs. \ref{eq:e0} and \ref{eq:a0}.
The binary forms at a true anomaly $f_f = 0.1 \pi$. We evolve this system 3 times: once with the original $e_0$ and $a_0$, and twice with
a slight alteration of the initial eccentricity that can arise from the tidal forces of the perturber. We decrease the initial eccentricity
by $10^{-4}$ and $10^{-3}$. The trajectories of each of these four systems are depicted in Fig. \ref{fig:traj_tides}.
In Fig. \ref{fig:dphi_tide}, we show their phase shift evolution as a function of time and peak GW frequency.

As seen, a small change in eccentricity may lead to drastic changes in the initial pericentre distance, especially in mergers that
form at such high eccentricities. The system that we study here has an initial eccentricity of $e_0 \sim 0.998$, and decreasing
this by only $10^{-3}$ would increase the pericentre distance by a factor of 1.49. This allows for less power emitted in GWs in these early stages,
therefore resulting in a longer inspiral time. With an eccentricity change of $10^{-4}$, the time to coalescence becomes $\sim1.17$
times larger. For the $10^{-3}$ case we obtain an increase in time as large as a factor $\sim$3.7. In terms of detectability, this might work in our favour.
However, as we see in the bottom panel of Fig. \ref{fig:dphi_tide}, this also has its downsides as it shifts the evolution of the phase shift to a lower peak GW frequency, potentially pushing it out of band. Given that
\begin{align}
    f_\text{p,GW} \sim \frac{1}{\pi} \sqrt{\frac{2Gm}{r_p^3}},
    \label{eq:fp_GW}
\end{align}
increasing $r_p$ by a factor 1.49 lowers the peak GW frequency at formation by a factor $\sim$1.8.

In conclusion, phase shifts may get largely amplified by tides from the perturber tidal effects, potentially increasing chances of a detection. Small drops in initial eccentricity lead to longer inspiral times and higher phase shifts as well as a shift to lower peak frequencies. The latter might result in the binary being pushed out of the detector band.

\section{Summary}\label{sec:summary}

In this paper, we explored for the first time the Rømer delay-induced phase shift, arising from an acceleration to the binary COM,
of an eccentric BBH that is on an eccentric orbit around a third object. Besides presenting results from a semi-analytical model
including orbital evolution using PN-orbit averaged equations and Kepler's Equations, we further presented several analytical approximations.
With these tools, we investigated the effects of the GW phase shift that are unique to the outer orbit being eccentric together with an eccentric evolving inner orbit merging BBH. Lastly, we applied our model to relevant astrophysical scenarios of chaotic 3-body scatterings, and studied the effect of tides from the third-body onto the GW phase shift.

The main takeaways from this work are the following:
\begin{itemize}
    \item For BBHs that inspiral along eccentric orbits the time evolution of the phase shift is dependent on where in the orbit the binary evolves.
    For low and moderate outer eccentricities, the leading order effect relates to how the magnitude of the GW phase shift can change over the orbit
    with the distance between the binary and third object, $r_m$, as $\propto r_{m}^{-2}$. This especially illustrates that
    the GW phase shift easily can change by order-of-magnitude over the orbit with its maximum around merger at pericentre.
    \item At high outer eccentricities ($e_\mathrm{out}=0.9$ in our example), effects of the strong curvature around pericentre are imprinted
    onto the phase shift. Two characteristics stand out: first, the maximum possible phase shift no longer occurs when the binary merges
    at pericentre, but slightly beyond ($f_m \sim -0.2\pi)$. Secondly, in certain parts of parameter space ($f_m \sim -0.75\pi$ in our example),
    two peaks may occur. If the binary passes pericentre before the main peak happens, a boost in phase shift due to the large curvature and
    high velocity may give rise to a secondary peak. These features are important properties of eccentric outer orbits..
    \item Chaotic scatterings are ideal environments for such phase shifts and their orbits have to be described by eccentric models, as the ones we
    presented here, compared to the standard circular.
    \item In the case of GW capture formation of BBHs at the Hill sphere, tides from the third body may give rise to small changes in the initial binary eccentricity. This induces an increased time to coalescence and thereby a higher maximum $d\phi$.
    At the same time, this pushes the peak frequency down, potentially out of band.
\end{itemize}

The completion of this work has opened up the window to numerous interesting future studies, including quantifying the detectability, GW phase statistics from cluster simulations, comparisons to other dynamical environments, as well as incorporating observer dependence
and other GR GW phase shift effects. We reserve that for future works. 

\section*{Acknowledgements}
It is our pleasure to thank Daniel J. D'Orazio, Martin Pessah, David W. O'Neill and Gaia Fabj for their extremely insightful comments and suggestions. K.H. is grateful for the fruitful discussion sessions at CIERA, Northwestern University, which helped complete this work. The authors are supported by Villum Fonden grant No. 29466, and by the ERC Starting Grant no. 101043143 — BlackHoleMergs.

\section*{Data Availability}
The data underlying this article will be shared upon reasonable request to the corresponding author.

\appendix
Below we link our model to actual GW observations by adding our phase shift to an eccentric waveform model in the LIGO band,
and investigate observable characteristics. As mentioned in Sec. \ref{sec:summary}, more detailed studies of the
detectability of the phase shift in GW signals will be executed in the near future. \newline

\section{Effect on gravitational waveform}

\begin{figure*}[t]
    \centering
    \subfloat{
    \includegraphics[width=.95\textwidth]{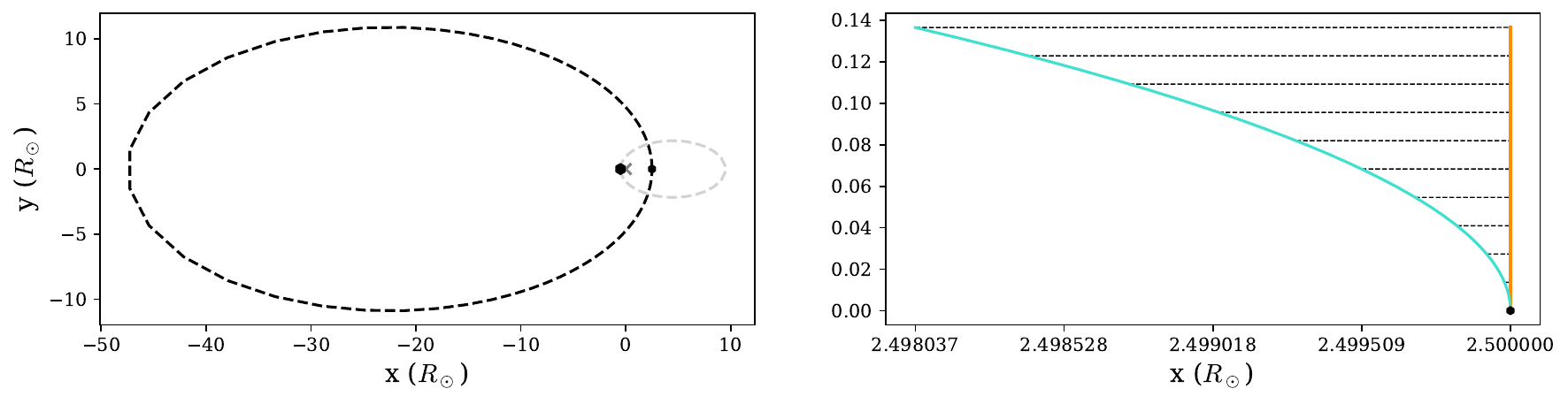}\label{fig:traj_ecc}
    } \\
    \subfloat{
    \includegraphics[width=.95\textwidth]{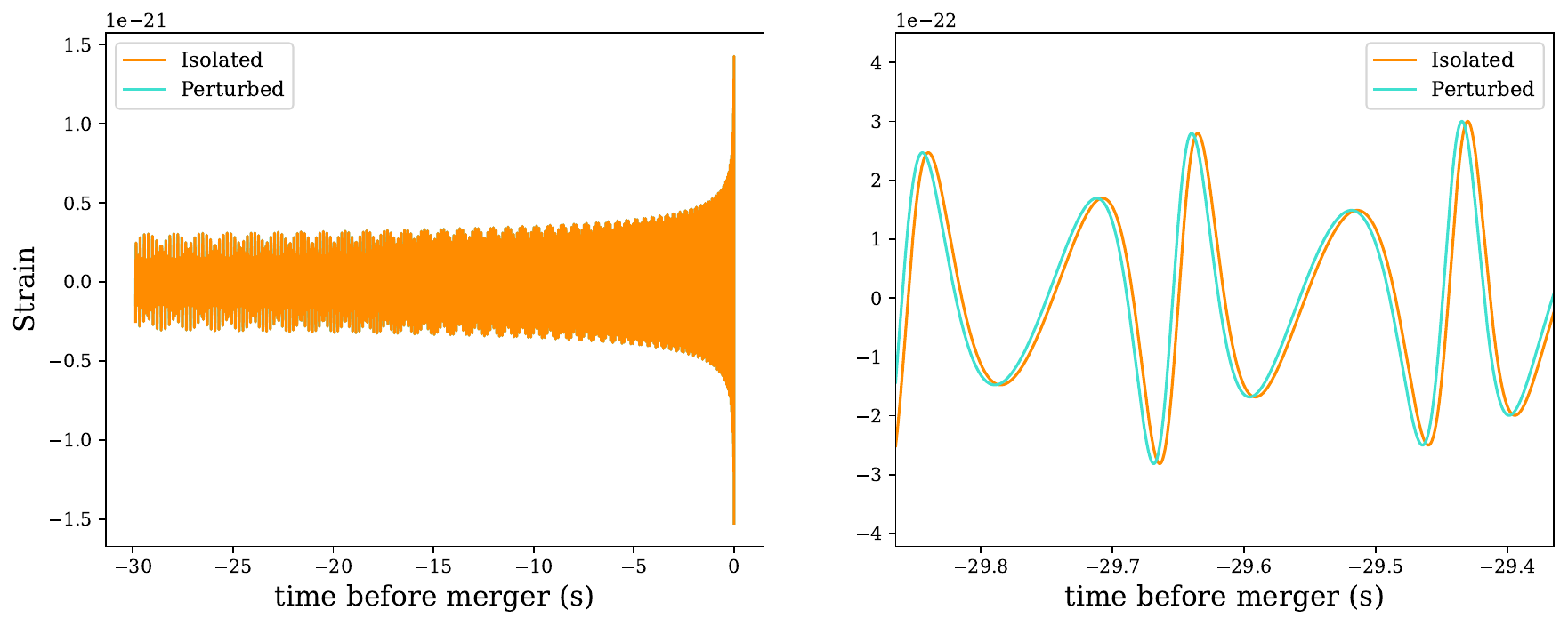}\label{fig:GW_ecc}
    } \\
    \subfloat{
    \includegraphics[width=.95\textwidth]{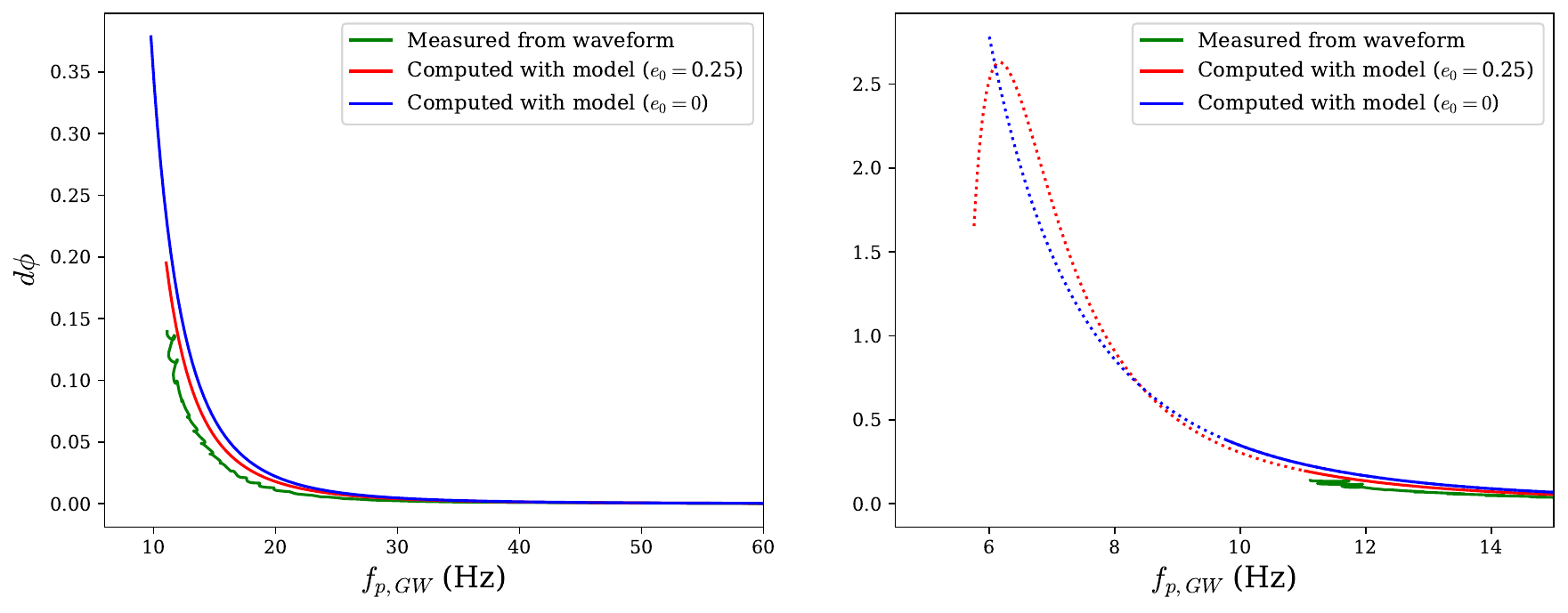}\label{fig:dphi_meas}
    }
    
    \caption{{\bf GW phase shift from real GW forms.} Binary parameters: $m_1 = m_2 = 5M_\odot$ with $e_0 = 0.25$ at $f_\text{GW}=10$ Hz, at a distance $D=100$pc and inclination $\iota=0$. For the triple, we use $m_3 = 100 M_\odot$, $a_\text{out} = 30 R_\odot$, and $e_\text{out} = 0.9$. {\textit{Top left}}: the trajectories of the binary COM (rightmost black circle) and perturber (leftmost black circle), around the triple COM (red cross). The dashed green auxilliary ellipse depicts the Keplerian outer orbit on which the binary COM is moving. The binary merges at the pericentre of the outer orbit ($f_m = 0$). {\textit{Top right}}: a zoomed-in version of the left panel, where we can distinguish the trajectory of the binary in the presence of a perturber (orange) and its isolated reference version with a constant speed equal to the orbital speed at merger (turquoise). The coloured lines represent the distance $l(f)$ in Eq. \ref{eq:def_deltat_RD} at different $t$ before merger. {\textit{Middle left}}: in orange, we show the eccentric gravitational waveform of the perturbed binary. Behind this, in turquoise, we show the waveform of the isolated binary. These two waveforms are shifted in time due to their associated Rømer delay. The waveforms were generated with the \texttt{EccentricTD} model of the \texttt{pycbc} package. {\textit{Middle right}}: zoomed-in version of the left panel. {\textit{Bottom left}}: in turquoise, we show the phase shift as a function of GW frequency of the perturbed binary, where we extract $\Delta t$, $T_{12}$, and $f_{GW}$ directly from the waveforms in the middle panels. The orange curve represents the phase shift of a binary with the abovementioned parameters, as computed by our semi-analytical model of Sec. \ref{sec:theory}. The purple curve shows the evolving phase shift for a binary with the same parameters, but one whose inspiral is entirely circular. {\textit{Bottom right:}} here, we show the same plot as on the left, but we extend our semi-analytical models down to a GW frequency of 6 Hz.}
    \label{fig:phase_shift_waveform}
\end{figure*}

The phase shift has a direct imprint on the gravitational waveform of the binary. In order to visualise this,
we depict the evolution of a perturbed eccentric binary including its waveform in Fig. \ref{fig:phase_shift_waveform}.
The parameters of the triple are listed in the caption.
The top two panels show the setup of the system and the
trajectories of the binary COM and the third object. Since we are tracking the evolution only from late inspiral up to
merger, the binary COM only covers a tiny part of the outer eccentric orbit, making the trajectory invisible in the top
left panel. Both paths are more visible in the zoomed-in version on the right. Note that the x- and y-axis are not on
the same scale, making the two trajectories look more different than they actually are. Using the \texttt{EccentricTD}
waveform model \citep{tanay_frequency_2016} in the \texttt{pycbc} Python package \citep{alex_nitz_gwastropycbc_2024},
we generated an eccentric waveform with $e_0=0.25$ at $f_\text{GW}=10$ Hz. This can be seen as a case study of an
eccentric binary in the LIGO band. This waveform is depicted in the middle panels of Fig. 
\ref{fig:phase_shift_waveform}. Here, two versions of this GW signal are seen; in orange we show the signal to which we
applied a Rømer delay corresponding to the triple parameters, and the turquoise curve represents its isolated reference
counterpart. In the left panel, it is very difficult to distinguish the isolated and perturbed signal. Their time-dependent offset is more prevalent on the right, where we show a zoomed-in window of the first few orbits after it
enters the detector band. This panel clearly highlights the presence of the time delay. At $f_\text{GW}=10$ Hz, this
setup yields a Rømer delay $\Delta t \sim 0.005$s.

We can track the evolution of the phase shift between these two waveforms as a function of the peak GW frequency
$f_\text{p,GW}$, i.e. the GW frequency that contains the most power (which is eccentricity-dependent).
In order to do so, we directly measure $\Delta t, T_\text{orb}$ and $f_\text{p,GW}$ from the waveform.
We plot the resulting phase shift in the bottom panels of Fig. \ref{fig:phase_shift_waveform} in turquoise. We compare
the directly measured phase shift to our semi-analytical model, which we plot in orange. All 7 initial parameters
necessary for our model are known, except for $a_0$; this elusive parameter was obtained from the initial binary
orbital period, measured from the waveform. It should be noted that our model returns the orbital frequency rather
than the (peak) GW frequency, so we must use Eq. \ref{eq:fp_GW}. In purple, we show the frequency evolution of the phase shift for a binary with the same parameters, but $e_0 = 0$.

We observe some fluctuations in the phase shift that we measured from the waveform. The binary is eccentric, so the GW
frequency shows oscillatory behaviour as it chirps due to the fact that the binary separation varies over the course of
an orbit. We can see that the eccentric and circular inspiral yield a clearly different phase shift. While they both
have similar shapes, the circular version grows much faster in this regime; at $f_\text{GW} \sim 10$ Hz, it is about 1.5
times larger than the eccentric case. At the same GW peak frequency, the orbital period of the eccentric binary is 
larger therefore resulting in a lower phase shift. The phase shift that we extracted from the waveform is slightly off
from both curves, but clearly fits the eccentric case better. The fact that the eccentric and circular case are 
potentially distinguishable at frequencies as high as $\sim$ 10 Hz is promising in the context of possible detections.
This difference becomes more apparent if we were to track the evolution down to lower frequencies. In the bottom right 
panel of Fig. \ref{fig:phase_shift_waveform}, we extend the phase shift down to a GW frequency of 6 Hz, which lies in 
the DECIGO band. The circular case continues to rise, while in the eccentric case the decreasing orbital period catches 
up with the increasing Rømer delay, resulting in the typical peak.

The phase shift that we extract from the waveform does not match perfectly with our eccentric model.
This is largely due to the discrepancy between Keplerian and PN descriptions of the parameters we extract.
The orbital parameters $e_0$ and $a_0$ (and therefore $T_\text{orb}$ as well) describe a closed Keplerian orbit,
but are not defined as such in the PN framework in which the gravitational waveform of Fig. 
\ref{fig:phase_shift_waveform} evolves \citep{damour_general_1985,memmesheimer_third_2004}. In other words, we are 
extracting Keplerian parameters from a system that is PN in nature. Additionally, we extract the peak GW frequency
directly by identifying the duration of the bursts in the waveform corresponding to pericentre passage. This somewhat
simplified picture may also give rise to extra errors between the model and measured $d\phi$-$f_\text{p,GW}$ relation. 
The fluctuations in the green curve arise from the presence of apsidal precession, which creates oscillations in the 
peak frequency. We have damped these oscillations using a moving-average filter in order to better isolate the
global evolution of the phase shift.

\newpage

\bibliographystyle{aasjournal}
\bibliography{main}

\end{document}